\documentclass[twocolumn,prb]{revtex4-1}
\usepackage{amssymb,amsmath,bm}
\usepackage{graphicx}%
\usepackage{dcolumn}
\usepackage{color}
\usepackage{longtable}
\usepackage[dvips]{epsfig}
\usepackage[perpage,symbol*]{footmisc}
\usepackage{multirow}
\usepackage{verbatim}
\usepackage{tablefootnote,threeparttable}
\usepackage{braket}
\usepackage{comment}
\usepackage{simpler-wick} 

\usepackage{lineno}

\usepackage[T1]{fontenc} 

\usepackage{filecontents} 

\renewcommand{\textcolor}[2]{#2}
\usepackage{xr}

\newcommand*{\citen}{}
\DeclareRobustCommand*{\citen}[1]{%
  \begingroup
    \romannumeral-`\x 
    \setcitestyle{numbers}%
    \cite{#1}%
  \endgroup
}

\newcolumntype{.}{D{.}{.}{-1}}
\newcommand{\mc}[3]{\multicolumn{#1}{#2}{#3}}

\newcommand{\fns}{\footnotesize}

\newcommand{\fig}[2]{\scalebox{#1}{\includegraphics{#2}}}

\usepackage{bbm}

\usepackage{hyperref}
\hypersetup{colorlinks,breaklinks,linkcolor=Blue,urlcolor=Blue,anchorcolor=Blue,citecolor=Blue}
\usepackage{color}
\definecolor{DarkBlue}{rgb}{0.0,0.08,0.45}
\definecolor{Blue}{rgb}{0.0,0.0,1.0}
\definecolor{Red}{rgb}{1.0,0.0,0.0}
\definecolor{RedOrange}{rgb}{0.9,0.0,0.2}
\definecolor{dgrn}{RGB}{0,150,0}
\definecolor{dgray}{gray}{0.3}

\makeatletter
\newcommand*{\addFileDependency}[1]{
  \typeout{(#1)}
  \@addtofilelist{#1}
  \IfFileExists{#1}{}{\typeout{No file #1.}}
}
\makeatother

\newcommand*{\myexternaldocument}[1]{%
    \externaldocument{#1}%
    \addFileDependency{#1.tex}%
    \addFileDependency{#1.aux}%
}

\myexternaldocument{2}

\usepackage{titlesec}  
\titleformat{\section}[hang]{\bfseries\large}{\thesection}{1em}{}
\renewcommand{\thesection}{\arabic{section}}
\renewcommand{\thesubsection}{.\arabic{subsection}}
\renewcommand{\thesubsubsection}{.\arabic{subsubsection}}
\titleformat{\section}{\normalfont\bfseries}{\thesection.~}{0.25em}{}
\titleformat{\subsection}[runin]{\normalfont\bfseries}{\thesection\thesubsection.}{0.25em}{}
\titleformat{\subsubsection}[runin]{\normalfont\bfseries}{\thesection\thesubsection\thesubsubsection.}{0.25em}{}
\begin{document}
\title{
Optimal-Reference Excited State Methods:
Static Correlation at Polynomial Cost with
Single-Reference Coupled-Cluster Approaches
}

\author{Sylvia J. Bintrim}

\author{
    Kevin Carter-Fenk
}
\email[Department of Chemistry, University of Pittsburgh, Pittsburgh, Pennsylvania 15218, USA; ]{kay.carter-fenk@pitt.edu}

\date{\today}


\begin{abstract}
\begin{center}
\textbf{\textcolor{red}{Abstract}}
\end{center}
Accurate yet efficient modeling of chemical systems with
pronounced static correlation in their excited states
remains a significant challenge in quantum chemistry, as most electronic structure methods that can adequately capture static correlation scale factorially
with system size.
Researchers are often left with no option but to use more affordable
methods that may lack the accuracy
required to model critical processes in photochemistry
such as photolysis, photocatalysis, and non-adiabatic relaxation.
A great deal of work has been dedicated to refining
single-reference descriptions of static correlation in the
ground state via
``addition-by-subtraction'' coupled cluster methods
such as pair coupled cluster with double substitutions
(pCCD), singlet-paired CCD (CCD0), triplet-paired CCD (CCD1),
and CCD with frozen singlet- or triplet-paired amplitudes (CCDf0/CCDf1).
By combining wave functions derived from
these methods
with the intermediate
state representation (ISR),
we gain insights into the extensibility of
single-reference coupled cluster theory's
coverage of static correlation to the excited state problem.
Our CCDf1-ISR(2) approach is robust in the face of
static correlation and provides enough dynamical
correlation to accurately predict excitation energies to
within about 0.2~eV in small organic molecules. We also highlight distinct advantages of the
Hermitian ISR construction, such as the avoidance of
pathological failures of equation-of-motion methods
for excited state potential energy surface topology.
Our results prompt us to continue exploring
optimal \textcolor{red}{single}-reference theories (excited state approaches
that leverage dependence on the initial reference
wave function) as a potentially economical approach
to the excited state static correlation problem.

\end{abstract}
\maketitle
\thispagestyle{plain}
\section{Introduction}
Electron correlation beyond Hartree-Fock theory
is the central problem in quantum chemistry.
While the distinction is somewhat arbitrary,
it is often conceptually useful to partition
the correlation energy into dynamical
and non-dynamical (static) correlation
effects.\cite{LeeTay89, BarSta94, MokNeuHan96,HanCoh01,BogTecLeg12, Cri13,TsuVan14,HolHosMen16,RamSalMat16,BenLatMar17,ViaRodRam19}
Of the two, dynamical correlation
is simpler to incorporate into theoretical
model chemistries as it manifests from spontaneous
repulsions between pairs of electrons that
can be incorporated into a single-reference formalism.
On the other hand, static correlation
results from many possible ground state reference determinants being of roughly equal energy and importance
, and its affordable incorporation into practical
calculations remains one of the grand
challenges of electronic structure
theory.

Density functional theory (DFT)
captures electron correlation at mean-field cost,
but the included correlation is almost exclusively dynamical,
and the Hohenburg-Kohn and Kohn-Sham theorems preclude any obvious extension of DFT to
the multi-determinant case,\cite{HohKoh64,KohSha65} though there
are efforts in this direction.\cite{Cre01,Cha12,Bec13b,SuLiYan18,YehYanHsu22}
Beyond the challenges posed by defining a static
correlation functional under the constraints of a
single, spin-pure determinant,
DFT methods are also afflicted by self-interaction
errors\cite{ZhaYan98b,MorCohYan06,MorCohYan08} that cause a myriad of problems including
underestimated barrier heights,\cite{PatZie02,ShuMisBar23}
spuriously low-energy
charge-transfer
excitations,\cite{TozAmoHan99,CasGutGua00,DreWeiHea03,DreHea04,BerSprHut04,NeuGriBae06,LanHer07,IsbMarCur13,MagTre07,PeaBenHel08,LanRohHer08,LiaFenHai22}
and fundamental difficulties
with local approximations innate to many-body expansion algorithms.\cite{BroHer24}

Unlike DFT, wave function theories (WFTs)
more naturally lend themselves
to a depiction of static correlation as
electron self-interactions can be exactly
eliminated.\cite{SzaOst82}
However, WFT methods that can treat static
correlation are few in number and generally
scale factorially with the number of correlated
orbitals in the system. Some notable examples are
complete active space (CAS) approaches wherein
the full configuration interaction (FCI) coefficients
are optimized, either alongside the molecular orbitals (MOs)
in a CAS self-consistent field (CASSCF) procedure
or without MO optimization
(CASCI).\cite{DasWah66,RooTaySig80,Roo80,SzaMulGid12}
While the extension to multireference situations remains
somewhat formally ambiguous, multireference
coupled-cluster (CC) theories are also a hotbed of active
methodological
development.\cite{Lin78,JezMon81,Eva18}
Furthermore, cumulant functional methods such as
density matrix functional\cite{Gil75,ZumMas85,Kut06,SokSch13,MenvanGri14,Sch18} and natural orbital
functional\cite{Mul84,GoeUmr98,PirOtt03,LeiPir05,RohPerGri08,Pir13,Pir14,Pir17}
theories alone,
or interfaced with WFT methods like
M{\o}ller-Plesset perturbation theory,\cite{Pir18,HolLoo20}
have also shown promise in economical descriptions of
static correlation.

The discussion heretofore has been dedicated exclusively to
electron correlation in the ground state because most efforts
towards describing static correlation have focused on improving the ground state.
When facing static correlation
problems in excited state calculations,
the computational
chemist's affordable options are sorely lacking.\cite{LisNacAqu18}
This deficiency in suitable options
can be problematic when modeling
photolysis or nonadiabatic dynamics more generally. For example, single-reference methods often overestimate nonradiative
decay yields in surface-hopping simulations.\cite{PapJacVac24}
Available multireference methods include CASCI,
state-specific
CASSCF,\cite{TraSheNeu19,TraNeu20,HanNeu22,TraNeu23,MarBur23}
state-specific CI,\cite{KosLoo23}
multiconfigurational linear response atop a CASSCF reference,\cite{YeaJor79,Dal80,OlsJor85}
excited state mean-field theory,\cite{ZhaNeu16,SheNeu18,HarNeu20,ZhaNeu20,SheGwiNeu20,CluSheNeu20}
and multireference algebraic diagrammatic construction (MR-ADC).\cite{Sok18,ChaSok19,ChaSok20,MazSok21,BanSok23}
Apart from excited state mean-field theory,
all of the aforementioned approaches use
a CASSCF reference state that requires a bespoke selection
of active orbitals, leading to varying results
depending on active space selection. In addition, the
factorial scaling with respect to active space size often limits
applications of CASSCF-reference theories to 
a \textcolor{red}{subset} of explicitly correlated orbitals,
possibly resulting in the omission of important correlation
effects. \textcolor{red}{Selecting an
active space guided by chemical intuition
alone amplifies this risk, but
the density matrix renormalization group (DMRG) can make larger active space sizes more accessible by application of singular value decomposition
to compress the variational wave function, and information-theoretic metrics can be used
to automatically select an active space.\cite{HuCha15, RenPenZha17, BaiRei20}}

Among the single-reference approaches that are capable of
accounting for some static correlation are spin-flip
variants of ADC,\cite{LefWorDre15,LefRehDre16,LefTunMar17} equation-of-motion CC (EOM-CC),\cite{KryShe02,LevKry04,Kry05}
and time-dependent DFT (TD-DFT),\cite{ShaHeaKry03,BerShaKry12} albeit at the cost of forgoing spin-pure excited states.
There has also been a recent surge of interest in seniority-zero
CC approaches such as the pair coupled cluster doubles (pCCD)\cite{SteHenScu14,HenBulSte14,BrzBogTec19,Bog21,Bar24}
theory (described below) \textcolor{red}{with extensions to excited states including EOM-pCCD, linear response pCCD, and orbital-optimized pCCD for doubly-excited states.}\cite{Bog16,Bog19,KosMarSce21,GayBog24,AhmBogTec24}
However, as we will later discuss, pCCD is not
invariant to unitary transformations of the occupied or virtual
orbitals. After orbital optimization or localization, the pCCD orbitals are
a well-defined basis for the pCCD energy, but the lack of orbital invariance
of the energy \textcolor{red}{may hinder} extensions to local correlation theories or fragment-based approaches
and may have a detrimental impact on excited state analyses.\cite{HerMan24} 
\textcolor{red}{Excited state properties may be adversely
affected by the fact that orbital-optimized pCCD introduces spurious spatial symmetry
breaking, which in concert with the orbital dependence of pCCD, may introduce
artifacts in the spectra of small, symmetric molecules such as broken degeneracies
and incorrect selection rules.}
\textcolor{red}{Despite the lack of orbital invariance, pCCD-based methods have found success in applications from
vertical excitation energies in actinide- and lanthanide-containing complexes
to modeling the structure of large organic molecules by interfacing
pCCD with embedding methods.\cite{NowTecBog19, TecBogBor19, TecGalSzc23}}

In this work, we present a polynomial-scaling,
black-box, single-reference, size-consistent excited state approach
that is robust in cases
of static correlation,
provides spin-pure excited states,
and employs a Hermitian approach necessary for
the description of excited state potential energy surface topology\cite{KohTaj07, KjoKoc17} and
for the size-intensivity of predicted oscillator strengths.
In particular, we will leverage
the sensitivity of perturbative excited state approaches
based on the intermediate state representation
(ISR)\cite{SchTro04,DreWor15,HodRehDre20,DrePapDem23,SchPapFra23}
to the initial reference wave function,
replacing the usual second-order M{\o}ller-Plesset
(MP2) reference wave function with the first-order
approximation to a CC wave function that captures
some static
correlation effects.
This approach was originally
introduced by Dreuw and co-workers using
CC with double substitutions
(CCD) or CC with single
and double substitutions (CCSD)
but has yet to be thoroughly explored for
other CC {\em ans{\"a}tze}.\cite{HodDemReh19,HodRehNor19,HodRehDre20}
Namely, ``addition-by-subtraction''
approximations wherein only
certain double substitutions are
retained ({\em i.e.} the
aforementioned pCCD, among others
discussed below)
have not been assessed in the ISR context.
Such approaches
might be better suited
for systems in which static correlation
is more pronounced,
thereby making the best of the
first-order CC wave function approximation within the ISR.\cite{Bar24}

At the heart of this work is the need to
provide computational chemists with more affordable
tools that can describe the many incidences 
in computational photochemistry where static correlation
becomes important, such as photolysis
and
the rich and useful
photochemistry of transition metal
complexes.\cite{ZobGon21}
The crux of our foray into such optimal \textcolor{red}{single}-reference ISR
approaches is the question, ``Does
the robustness of addition-by-subtraction CC theory
to static correlation translate to excited states?''
Herein we demonstrate
the potential power of polynomial-scaling ISR approaches
for capturing a wide array of crucial properties in
the photochemistry of statically-correlated systems
without compromising the
accuracy of ISR methods in weakly-correlated
systems where the performance of ADC
is already satisfactory.

\section{Theory}
\subsection{Addition-by-Subtraction CCD}
The standard CCD approach employs an exponential
{\em ansatz} to the wave function,
\begin{equation}
|\Psi_{\text{CC}}\rangle = e^{\hat{T}_2}|\Phi_0\rangle\;~,
\end{equation}
where $|\Phi_0\rangle$ is usually
the Hartree-Fock ground state reference determinant,
\begin{equation}
    \hat{T}_2 = \frac{1}{4}\sum\limits_{ijab}t_{ij}^{ab}\hat{a}_a^\dagger\hat{a}_b^\dagger\hat{a}_j\hat{a}_i
\end{equation}
is the usual double-substitution operator,
and $\hat{a}_i$ and $\hat{a}_a^\dagger$ are
particle annihilation and particle creation
operators, respectively.
Throughout this work, occupied
orbitals will be indexed as $\{i,j,k,l,\dots\}$
and virtual orbitals as $\{a,b,c,d,\dots\}$.
The corresponding energy and amplitude equations
for CCD are
\begin{subequations}\label{eqn:main}
\begin{alignat}{1}
\langle\Phi_0|e^{-\hat{T}_2}\hat{H}e^{\hat{T}_2}|\Phi_0\rangle  &= E\label{subeqn:a}\\
\bra{\Phi_{ij}^{ab}}e^{-\hat{T}_2}\hat{H}e^{\hat{T}_2}\ket{\Phi_0}&=0
\label{subeqn:b}
\end{alignat}
\end{subequations}

Other authors have shown quite convincingly
that judicious removal of some $\hat{T}_2$
amplitudes can greatly improve the qualitative
behavior of CCD when static correlation
becomes more important.\cite{PiePal91, KatMan13, BulHenScu15, Pal17, Bar24}
The most aggressive such
approximation is known as pCCD,
wherein
all but the diagonal components of
$\hat{T}_2$ are removed, resulting in
\begin{equation}
    \hat{T}_2 = \sum\limits_{i\bar{i}a\bar{a}}t_{i\bar{i}}^{a\bar{a}}\hat{a}_a^\dagger\hat{a}_{\bar{a}}^\dagger\hat{a}_{\bar{i}}\hat{a}_i
\end{equation}
where barred indices correspond to $\beta$-spin orbitals.
The pCCD approach is equivalent to 
a generalized valence bond method known as the
antisymmetric product of 1-reference orbital geminals
and describes single- \textcolor{red}{ and double-}bond breaking quite
well.\cite{LimAyeJoh13,TecBogJoh14,BogTecAye14,BogTecLim14,BogTecBul14,LimKimAye14, LesMatLeg22}
Combined with its $\mathcal{O}(N^3)$ computational
complexity (after one $\mathcal{O}(N^5)$ transformation
from the atomic orbital to molecular orbital basis),
the robustness of pCCD
in the face of bond breaking has driven considerable
interest despite the fact that 
pCCD lacks invariance to orbital rotations
within the occupied or virtual subspaces.

Orbital invariant approaches that retain only the
singlet (CCD0) or triplet (CCD1) amplitudes
have shown similar success to pCCD in the description
of bond breaking, albeit at an increased cost
of $\mathcal{O}(N^6)$.\cite{BulHenScu15,GomHenScu16} 
Singlet-paired CCD0 substitutions take the form
\begin{equation}
    \hat{T}_2^{[0]} = \frac{1}{2}\sum\limits_{ijab}\sigma_{ij}^{ab}P_{ab}^\dagger P_{ij}
\end{equation}
where
\begin{equation}\label{eqn:SingletOperators}
    P_{ij} = \frac{1}{\sqrt{2}}(\hat{a}_j\hat{a}_{\bar{i}}+\hat{a}_i\hat{a}_{\bar{j}})
\end{equation}
with the corresponding definition
for the singlet-paired creation
operator, $P_{ab}^\dagger$.
The singlet amplitudes are related to
the usual CCD $t_{ij}^{ab}$ by
\begin{equation}
    \sigma_{ij}^{ab} = \frac{t_{ij}^{ab}+t_{ij}^{ba}}{2}
\end{equation}
Likewise, triplet-paired CCD1 takes the form
\begin{equation}\label{eqn:T2_Triplet}
    T_2^{[1]} = \frac{1}{2}\sum\limits_{ijab}\pi_{ij}^{ab}\vec{Q}_{ab}^\dagger\vec{Q}_{ij}
\end{equation}
The product between
vector operators $\vec{Q}_{ab}^\dagger\cdot\vec{Q}_{ij}$
is evaluated as
\begin{equation}
    \vec{Q}_{ab}^\dagger\vec{Q}_{ij} = \big(Q_{ab}^+\big)^\dagger Q_{ij}^+ + \big(Q_{ab}^-\big)^\dagger Q_{ij}^- + \big(Q_{ab}^0\big)^\dagger Q_{ij}^0
\end{equation}
where their components are (again showing
only the annihilation operators explicitly)
\begin{subequations}\label{eqn:TripletOperators}
    \begin{alignat}{1}
        Q_{ij}^+&=\hat{a}_j\hat{a}_i\label{subeqn:TO_a}\\
        Q_{ij}^- &= \hat{a}_{\bar{j}}\hat{a}_{\bar{i}}\label{subeqn:TO_b}\\
        Q_{ij}^0 &= \frac{1}{\sqrt{2}}(\hat{a}_j\hat{a}_{\bar{i}}-\hat{a}_i\hat{a}_{\bar{j}})\label{subeqn:TO_c}
    \end{alignat}
\end{subequations}
Finally, the $\pi_{ij}^{ab}$ amplitudes are
related to the usual CCD amplitudes by
\begin{equation}
    \pi_{ij}^{ab} = \frac{t_{ij}^{ab} - t_{ij}^{ba}}{2}
\end{equation}
These singlet- and triplet-paired amplitudes are
uncoupled, resulting in an unsatisfactory recovery
of dynamical correlation in CCD0 and CCD1.
In this work, we will explore the uncoupled CCD0
and CCD1 reference wave functions alongside the
FpiCCD approach where the $\pi$ amplitudes
are computed, then frozen and inserted into the full
CCD equations.\cite{GomHenScu16} Hereafter we call this recoupling approach CCDf1, where ``f1'' means that the triplet
amplitudes are frozen. We refer to the analogous method where the singlet amplitudes are computed first and then frozen as CCDf0.

Both CCDf1 and CCDf0 can be viewed as infinite-order solutions to external
CC perturbation theory (xCCPT) equations,
where $\hat{T} = \hat{T}_x + \delta\hat{T}$
are the full $\hat{T}$ amplitudes: a set of external (or frozen)
$\hat{T}_x$ and a perturbation $\delta\hat{T}$.\cite{LotBar11}
In the case of CCDf1, we take $\hat{T} = \hat{T}_2$,
and $\hat{T}_x = \hat{T}_2^{[1]}$ are the frozen triplet
amplitudes. The wave function can be written as
\begin{equation}
    |\Psi_{\text{CC}}\rangle = e^{(\hat{T}_2^{[1]}+\delta\hat{T})}|\Phi_0\rangle
\end{equation}
resulting in the Schr{\"o}dinger equation
\begin{equation}
    \hat{H}e^{(\hat{T}_2^{[1]}+\delta\hat{T})}|\Phi_0\rangle = Ee^{(\hat{T}_2^{[1]}+\delta\hat{T})}|\Phi_0\rangle
\end{equation}
Pre-multiplying by $e^{-\hat{T}_2^{[1]}}$ gives
\begin{equation}\label{eq:xCC}
    \hat{X}e^{\delta\hat{T}}|\Phi_0\rangle = Ee^{\delta\hat{T}}|\Phi_0\rangle
\end{equation}
where $\hat{X}=e^{-\hat{T}_2^{[1]}}\hat{H}e^{\hat{T}_2^{[1]}}$ is the
once-similarity-transformed Hamiltonian.
This equation can be solved perturbatively, or (in the case of
CCDf1 and CCDf0) exactly for $\delta\hat{T}$ and $\delta E$.
In the latter case, we note that this corresponds to an infinite-order
solution to the external CC equations.
\subsection{Intermediate State Representation}
Within the ISR approach, excitation energies
are obtained by solving the
secular equation of a shifted
Hamiltonian $\hat{H}-E_0$,
\begin{equation}\label{eqn:ISR}
M_{IJ}=\langle\tilde{\Psi}_I|\hat{H}-E_0|\tilde{\Psi}_J\rangle
\end{equation}
which presents as a
Hermitian
eigenvalue problem,
\begin{equation} \mathbf{MX}=\mathbf{X}\bm{\Omega},\qquad \mathbf{X}^\dagger\mathbf{X}=\bm{1}
\end{equation}
Key to the ISR approach is the correlated
excited state basis,
\begin{equation}
    |\Psi_J^0\rangle = \hat{C}_J|\Psi_0\rangle
\end{equation}
where
\begin{equation}
    \hat{C}_J = \{a_a^\dagger a_i; a_a^\dagger a_i a_b^\dagger a_j, a<b, i<j;\dots \}
\end{equation}
are physical excitation operators that generate
the excited state configuration $J$.
The correlated excited states $\{\Psi_J^0\}$ are then
orthogonalized via the Gram-Schmidt procedure
to all intermediate states of
lower excitation classes and
then orthonormalized amongst themselves
to generate the intermediate states
$\{\tilde{\Psi}_J\}$.

The ISR procedure produces excited states
that are properly orthogonal to one another and
to the reference state, and the matrix
$\mathbf{M}$ is Hermitian by construction,
making ISR a convenient formalism to obtain
excited states and size-intensive properties.
Notably, the ISR formalism expands the solutions
to Eq.~\ref{eqn:ISR} order-by-order both
in the fluctuation potential and in terms
of the physical excitation operators.
For example, first-order in the fluctuation
potential corresponds to single excitations
only, while second-order in the fluctuation
potential results in a second-order
treatment of the single excitations with
a zero-order treatment
of double excitations, and so on.

\subsection{CC-ISR(2) Approximations}
As with ground state perturbation theory,
second-order ISR [ISR(2)] energies are
obtained using first-order wave functions
of the form
\begin{equation}
\label{eqn:ISR_wfn}
    |\Psi\rangle = |\Phi_0\rangle + |\Psi^{(1)}\rangle = (1+\hat{T}_2)|\Phi_0\rangle
\end{equation}
Dreuw and co-workers noted that, while not
strictly formally justifiable, the first-order
Taylor-expanded coupled cluster wave function can be written as\cite{HodRehDre20}
\begin{equation}
    e^{\hat{T}_2}|\Phi_0\rangle\approx (1+\hat{T}_2)|\Phi_0\rangle
\end{equation}
and inserted directly into the second-order
ISR(2) procedure to obtain an approach
that they called CCD-ISR(2).
The CCD-ISR(2) equations (Eq.~36 in Ref.~\citen{HodRehDre20})
differ from the
simplified ADC(2) equation because the CC $t$
amplitudes differ from those of MP2
(the typical choice for the ADC(2) ground state).
In preliminary studies of the CC-ADC(2)
{\em ansatz} (the direct use of the simplified
ADC(2) expression with CC amplitudes),
it was shown that CC amplitudes can improve
singlet-to-triplet excitation energies
and may be more robust against the divergences
that plague the MP2 reference.\cite{HodDemReh19,HodRehNor19}

Although the first-order approximation
to the CCD reference wave function is generally
stable in bond-breaking problems, CCD
struggles to dissociate multiple bonds
and may perform poorly in systems where
static correlation is important.\cite{BulHenScu15}
Inspired by
the relatively positive results
obtained using CC-ISR(2), we are interested in further exploring the
potential scope of Hermitian excited state methods
that result from the ISR construction.
In this work, we will explore the use
of addition-by-subtraction
CC amplitudes within the CC-ISR(2) framework.
Specifically, we will insert
pCCD, CCD0\slash1, and CCDf0\slash1 amplitudes
into Eq.~\ref{eqn:ISR_wfn}
to explore whether the qualitatively good
behavior of these approaches
for statically-correlated ground states
are translatable into the excited state manifold
via ISR(2).
\subsection{Brueckner orbitals}
Brueckner orbitals result from rotating the HF determinant to a basis in which the CC $t_1$ amplitudes are all zero. 
In the Brueckner orbital basis, there is
some account of orbital relaxation through inclusion of $\hat{T}_1$ substitutions
by nature of Thouless' Theorem.\cite{Tho60}
One consequence of
including additional relaxation effects
is that Brueckner orbitals often
preserve spatial symmetry, even when the HF
determinant breaks it.\cite{StaGauBar92}

By setting $t_1$ to 0 in the CCSD equations, we obtain the occupied-virtual block of the Brueckner modified Fock matrix:
\begin{equation}
\begin{split}
F_{ia} &= f_{ia}+\sum_{jb}f_j^bt_{ij}^{ab}+\sum_{jbc}\lbrack 2(ab|jc)-(ac|jb)\rbrack t_{ij}^{bc}
\\&-\sum_{jkb}\lbrack 2(ij|bk)-(ik|bj)\rbrack t_{jk}^{ab}
\end{split}
\end{equation}
in terms of spatial orbitals. 
Since the mean field self-consistency condition is $F_{ia}=0$, we 
can use existing SCF machinery to successively diagonalize
the re-defined Fock matrix to obtain a set of Brueckner orbitals.
Alternatively, one can directly
apply rotations via $e^{\hat{T}_1}$ to the MO coefficients until $t_1=0$
and then semi-canonicalize the orbitals. We take the latter approach in this work. 

The case of Brueckner pCCD
(pBCCD) is notably different
due to the lack of orbital
invariance of the pCCD energy.
Specifically, pBCCD
is often accompanied by successive
orbital localizations between Brueckner cycles to impart orbital
rotations that more significantly incorporate relaxation
into the MO basis.
Furthermore, Brueckner orbitals without localization do not provide a unique orbital pairing scheme for pCCD.\cite{SteHenScu14} 
Orbital localization results
in \textcolor{red}{artificial}
spatial symmetry breaking that
breaks important
degeneracies. \textcolor{red}{Because
pCCD is sensitive
to these changes, localized orbitals may
make pCCD excited states difficult to assign
and
influence selection rules that are
leveraged by spectroscopists to inform on
{\em physical}
symmetry-breaking in the active sites of metalloenzymes,
for instance.}\cite{SarAboFuj06}
\textcolor{red}{We leave detailed study of pCCD oscillator strengths for future work, but for the
aforementioned reasons,} we chose to use semi-canonical pBCCD/pBCCD-ISR(2) without
successive orbital localization.

In some sense, Brueckner CCD (BCCD)-ISR(2)
is more formally justifiable than CCD-ISR(2), as the $t_1$ amplitudes
are zero, akin to the usual MP2/ADC(2) case.
This formally decouples
single excitations from the reference
determinant in the CI-like ISR(2)
Hamiltonian, whereas
CCD-ISR(2) methods simply neglect this
coupling.
Formally speaking, methods such as CCD- and CCSD-ISR(2) have nonzero $t_1$ contributions that should impact the
singles block of the ISR matrix at
all orders in perturbation theory,
but these contributions are typically
ignored.
Herein, we compare the quality of ISR(2) results derived from CC and BCC ground states by neglecting the
$t_1$ contribution to ISR in the first
(usual) case
and formally decoupling the reference
state from the single excitations
in the latter.


\section{Computational Details}
The 
potential energy surfaces 
of N$_{\text{2}}$ \textcolor{red}{were} calculated
using
CASSCF with \textcolor{red}{second-order $N$-electron valence state perturbation theory} (CASSCF@\textcolor{red}{NEV}PT2) \textcolor{red}{using 6 active electrons and 6 active orbitals in the aug-cc-pVTZ basis set,\cite{Dun89} with resolution of the identity and chain of spheres exchange\cite{HeldeSNee21} for electron-repulsion integrals. For both ground and excited state potential energy surfaces, CASSCF orbitals optimized for the ground state were used.}
Time-dependent DFT (TD-DFT)
calculations used to generate potential energy
surfaces for formaldehyde made use of the
$\omega$B97X-D functional and
aug-cc-pVDZ basis set,\cite{ChaHea08b}
because according to recent benchmarks, $\omega$B97X-D is one of the best functionals in general for TD-DFT calculations on small organic molecules.\cite{LiaFenHai22}
Furthermore, the TD-DFT calculations feature a fairly dense
quadrature grid using 99 radial and 590 angular points
per atom for the integration of the exchange-correlation
potential.\cite{Leb75a,Leb75b}
All CC-ISR(2), ADC(2), and EOM-CCSD calculations on
N$_{\text{2}}$ and the Quest~\#1 database\cite{LooSceBlo18}
use the aug-cc-pVTZ basis set. \textcolor{red}{Calculations reported for the alkenes used the def2-TZVPP basis set, and basis set dependence in these systems was assessed using def2-SVP, def2-TZVP, and def2-QZVP basis sets.}\cite{Wei06} The oo-pCCD, EOM-pCCD+S, and EOM-oo-pCCD+S calculations on N$_{\text{2}}$ used the cc-pVDZ basis. 
We employed the aug-cc-pVDZ basis for
all calculations of
the formaldehyde potential energy surface.

The TD-DFT calculations presented herein were
carried out using Q-Chem~v6.2.\cite{QCHEM5} 
\textcolor{red}{CASSCF(6,6)@NEVPT2 calculations were performed using ORCA~v5.0.\cite{ORCA5}}
oo-pCCD, EOM-pCCD+S, and EOM-oo-pCCD+S calculations were performed using PyBEST~v2.0.0.\cite{BogLesNow21, BogBrzCha24}
Otherwise, all calculations were performed in a locally modified version of the PySCF software package.\cite{PySCF1,PySCF2, StaSok24}

\section{Results \& Discussion}
\subsection{\textcolor{red}{Hubbard Model}}
We begin our assessment of CC-ISR(2) approaches
by plotting the ground state and first
singlet-excited state of a ten-site, one-dimensional Hubbard
model at half-filling as a function of interaction strength ($U/|t|$). \textcolor{red}{The Hubbard excitations include contributions from double and higher excitation components that become more important as a function of interaction strength (Table~\ref{tab:SI-Hubbard-CI}).
While our CC-ISR(2) methods are single-reference, the shifted ISR Hamiltonian does contain a
(zeroth-order) description of double-excitations, so it may be possible that CC-ISR(2) methods can qualitatively describe
such collective excitations to a certain extent.
Our objective here is to assess (1) when the excited state character becomes too collective to be described
by our single-reference methods
and (2) how much the ground-state reference amplitudes matter in such a description.}

\begin{figure}[h!!]
    \centering
    \fig{0.5}{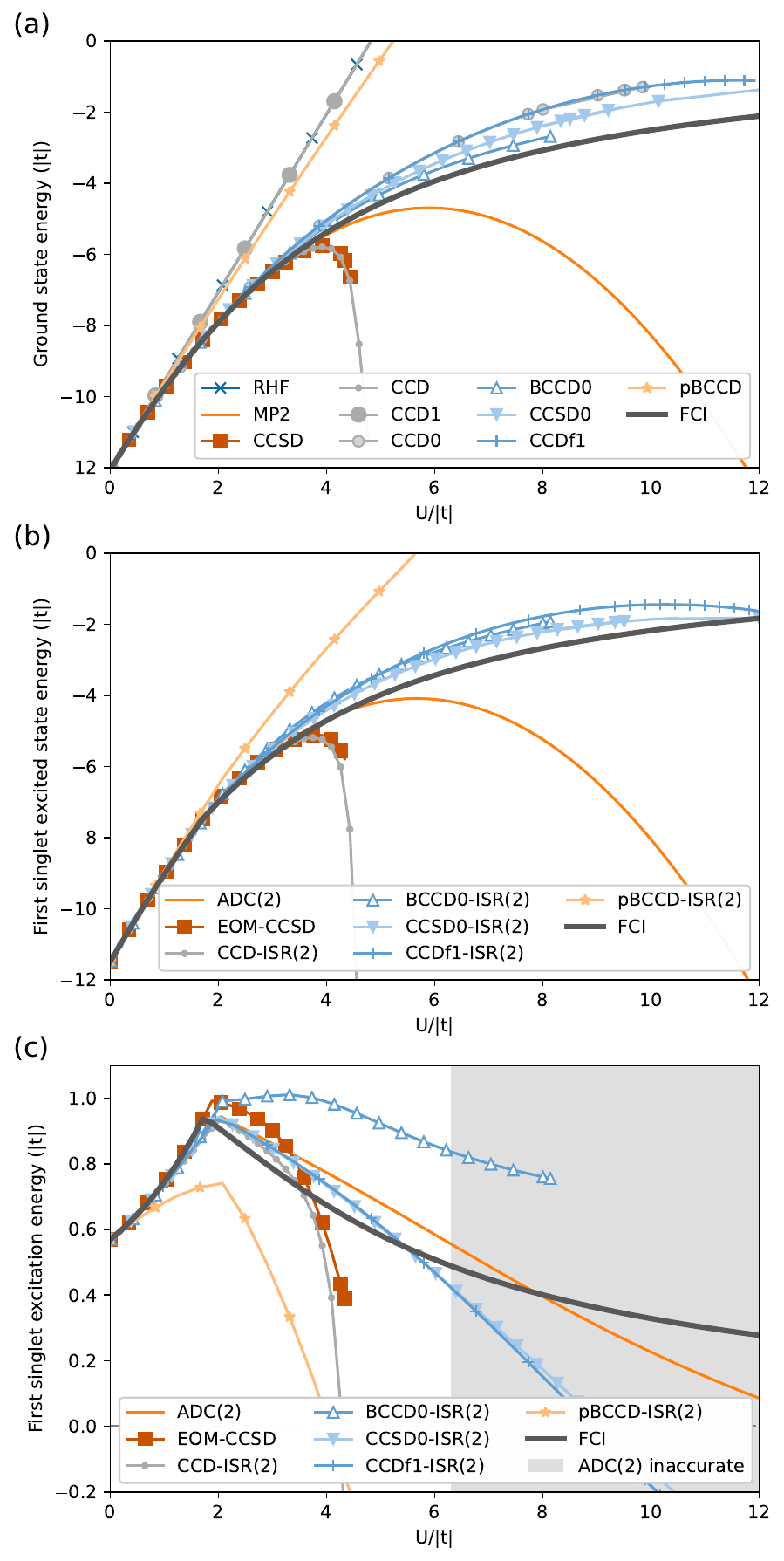}
    \caption{
    (a) Ground state ($E_{\text{S}_0}$),
    (b) first singlet excited state ($E_{\text{S}_1}$),
    and (c) first singlet
    excitation ($E_{\text{S}_1} - E_{\text{S}_0}$) energies
    as a function of interaction
    strength ($U/|t=-1.5|$) for a 10-site,
    half-filled Hubbard model
    with open boundary conditions.
    The FCI result is exact for both
    states and acts as the reference.
    The CC\slash CC-ISR(2),
    CCSD\slash EOM-CCSD, and
    MP2\slash ADC(2) results
    are computed using canonical Hartree-Fock
    orbitals, and the BCC \slash BCC-ISR(2) results use Brueckner orbitals. 
    The gray shaded area in (c) indicates
    where the MP2 energy from (a) begins to
    decrease, signaling divergence of
    the reference energy.}
    \label{fig:Hubbard}
\end{figure}

The results in Fig.~\ref{fig:Hubbard}a
are as-expected for the
ground state of the Hubbard
model, as each of these methods (apart
from CCDf1) have been assessed before for the Hubbard model.\cite{SteHenScu14, BulHenScu15, KelTsaReu22, CovTew23}
The CCD and CCSD energies diverge around
$U/|t| = 4$ while MP2 diverges
slightly later, monotonically decreasing
with interaction strength after $U/|t| \sim 6$.
With canonical Hartree-Fock (HF) 
or even Brueckner orbitals, pCCD fails
to capture much correlation. 
This linear increase in the pCCD
energy can be amended by
using orbitals
optimized with the pCCD Lagrangian (Fig.~\ref{fig:Hubbard_pccd_SD}), \emph{localized} Brueckner orbitals, or frozen pair CCD.\cite{SteHenScu14, HenBulScu15}
Unlike the other approaches,
the CCD0, BCCD0, CCSD0, and CCDf1 ground state energies are qualitatively
consistent with the exact (FCI) result. 
We note that the absence of off-site interactions within this Hubbard model
causes CCD0 and CCDf1 to give the same result,
as CCD1 adds no additional correlation energy to the HF
result under these conditions. BCCD0 falls slightly closer than CCSD0 to the FCI curve, and both perform better than CCD0, suggesting that inclusion of $\hat{T}_1$ can be helpful for quantitative accuracy. 

The lowest-energy singlet excited state
of the Hubbard model is shown in Fig.~\ref{fig:Hubbard}b and is qualitatively
similar to the ground state case for most methods.
The CCD-ISR(2) and EOM-CCSD approaches
once again diverge at
$U/|t|\sim 4$, suggesting
that they are not very well suited
for cases of substantial static correlation.
ADC(2) diverges in a way that is very similar
to its MP2 reference, which is to be expected.

However, there are some intriguing differences in
the Hubbard excited state
energies.
With canonical HF orbitals, pCCD-ISR(2) yields divergent Hubbard excitations closely tracking those of ADC(2), as shown
in Fig.~\ref{fig:Hubbard_pccd_SD}b.
We emphasize that our pCCD-ISR(2) matrix included the full singles-doubles block coupling, i.e. all double physical excitation operators were used to generate the intermediate states. In an effort to remedy the failure of canonical orbital pCCD-ISR(2), we first removed the non-pair double excitations from our ISR(2) matrix, opting
instead to use only the zeroth-order pair double excitations that are parameterized in
the pCCD Hamiltonian.
This is equivalent to solving the pCCD-ISR(2) problem restricted to the singles configuration space, as pair double excitations do not couple
to single excitations at first order.
With this modification, the pCCD-ISR(2) excited state still diverged. 
In contrast, we found that EOM-pCCD+S (even without optimized orbitals) predicts a non-divergent Hubbard excited state,
inspiring us to investigate the differences between EOM-pCCD+S and pCCD-ISR(2) more deeply.

One key difference between EOM-pCCD+S and pCCD-ISR(2) is that the former includes couplings
between the reference ground state and the singly-excited determinants.
To study the potential impact of reference/singles
coupling, we rigorously eliminate the coupling between the single excitations
and the ground state by employing a pBCCD reference without localization,
\textcolor{red}{so that the reference MOs obey
molecular point-group symmetry, akin to} the conditions in our EOM-pCCD+S calculations.\cite{SteHenScu14}
Fig.~\ref{fig:Hubbard_pccd_SD} shows that
the resultant pBCCD-ISR(2) gives non-divergent
Hubbard excitations similar to EOM-pCCD+S,
emphasizing the sensitivity of methods that lack orbital invariance to orbital
rotations (even when said rotations are supplied in the CI-like EOM Hamiltonian).
While there is no reason to expect ISR and EOM
approaches to agree quantitatively, we find
it enlightening that decoupling the single excitations
from the reference state leads to such similar
results, essentially implicating the $\hat{T}_1$
contribution for the qualitative differences.

In both ground and excited states, inclusion of $\hat{T}_1$ in the ground state reference improves the results over those of CCD0/CCD0-ISR(2). Although BCCD0 performs better than CCSD0 in the ground state, the reverse is true in the excited state.  We explain below that this is likely caused by differences in how CC and CC-based ISR(2) treat the ground state wavefunction along with further approximations
that neglect $t_1$ contributions in the CCSD0-ISR(2)
treatment.

Overall, the
methods that are most robust as
the interaction strength increases are
CCD0-ISR(2) and CCDf1-ISR(2), with or without the inclusion of $\hat{T}_1$ -- although their excited state energies begin (incorrectly) to decrease after $U/|t|\sim 10$. \textcolor{red}{Many physical systems fall within the range $ U/|t|\leq 8$, with $U/|t|\geq 4$ considered strong correlation,\cite{DelGingHol09, QinSchAnd22, MooHorLin24} so} 
we anticipate rather robust treatment\cite{DelGingHol09, QinSchAnd22, MooHorLin24}
of strongly correlated systems within
CCD0-ISR(2) and CCDf1-ISR(2). \textcolor{red}{Interestingly, CCD0-ISR(2) performs better than EOM-CCSD0, if we judge the methods' performance by the interaction strength at which the excited state energies begin to incorrectly turn over (Fig.~\ref{fig:Hubbard_eom_ccsd0}).}

We can gain even more insight into the
CC-ISR(2) approach by examining the excitation energies within the Hubbard model. The results in Fig.~\ref{fig:Hubbard}c show the
energy difference between the first singlet excited state and the
ground state. The exact result
exhibits a maximum near $U/|t|=2$ and
then decreases monotonically. 
The CCD-ISR(2) and EOM-CCSD approaches
fail quickly with interaction strength,
as suggested by the previous results.
Perhaps counterintuitively, despite
the divergent behavior of MP2 and ADC(2)
in the absolute energies of the two states,
the energy difference
yields reasonable results well beyond
the point at which MP2 diverges.

Despite
the relatively good performance of
CCD0-ISR(2), CCSD0-ISR(2), and CCDf1-ISR(2) in terms
of total energies, the excitation energies
begin to deviate from FCI at interaction strengths beyond
$U/|t| \sim 6$.
We attribute this to an imbalance between
the treatment of the ground state
wave function in CC and the excited state wave function in the ISR(2) procedure. For example, in CCD0, the ground state wave function is written as $e^{\hat{T}_2^{[0]}}|\Phi_0\rangle$, while the ground state wave function is approximated as $(1+\hat{T}_2^{[0]})|\Phi_0\rangle$ in the ISR(2) procedure. Whereas the CC ground state wave function
and energy are calculated using
the full exponential operator, thus
folding in effects of quadratic
contributions $\hat{T}_2^2$,
the excited state ISR(2) wave function
is built by applying physical excitation operators to a first-order approximation to the exact ground state wave function.
ADC(2) does not experience such an imbalance,
as the ground- and excited state wave functions
are both treated to first order;
so while they both diverge, they do so
at similar rates, such that the energy
differences remain misleadingly reasonable.
In the case of CCD0-ISR(2), the different treatment
of the ground- and excited state wave functions
results in an over-stabilization of the
excited state as the interaction
parameter becomes very large. 

Unlike CCSD0-ISR(2), BCCD0-ISR(2) over-estimates absolute excitation energies. The Brueckner mean field determinant produces a CC wavefunction with rigorously zero contribution from $\hat{T}_1$, so that the energy of the ISR(2) ground state wavefunction $(1+\hat{T}_2)\ket{\Phi_0}$ is less under-estimated. Combined with the fact that the BCCD0 energy is lower than that of CCSD0, we obtain BCCD0-ISR(2) excitation energies that are much larger than those of CCSD0-ISR(2). Surprisingly, despite the significant overestimation
of excitation energies, BCCD0-ISR(2) appears to
qualitatively improve the asymptotic behavior of the excitation
energy surface at larger interaction strengths.
This implies that using Brueckner orbitals instead of canonical orbitals may be better
suited for the qualitative description of static correlation
in excitation energy differences.


While excitation energies are
certainly the most common
metric for evaluating the performance
of excited state methods,
our results from the Hubbard model
suggest that total energies 
should also be
considered as important measures of
a method's accuracy. This is especially
important in the case of ADC(2), where
the excitation energies are qualitatively
reasonable, but the forces ($-\nabla V$)
in the ground and excited states are surely
incorrect. \textcolor{red}{Beyond energetic comparisons, correlation spectra and orbital-pair mutual information diagrams would be helpful to further
discern which methods better capture the actual Hubbard model physics, but we leave this deeper analysis to future work.\cite{BogTecLeg16, NowLegBog21}}
\subsection{\textcolor{red}{N$_2$ Dissociation}}
\begin{figure*}
    \centering
    \fig{0.6}{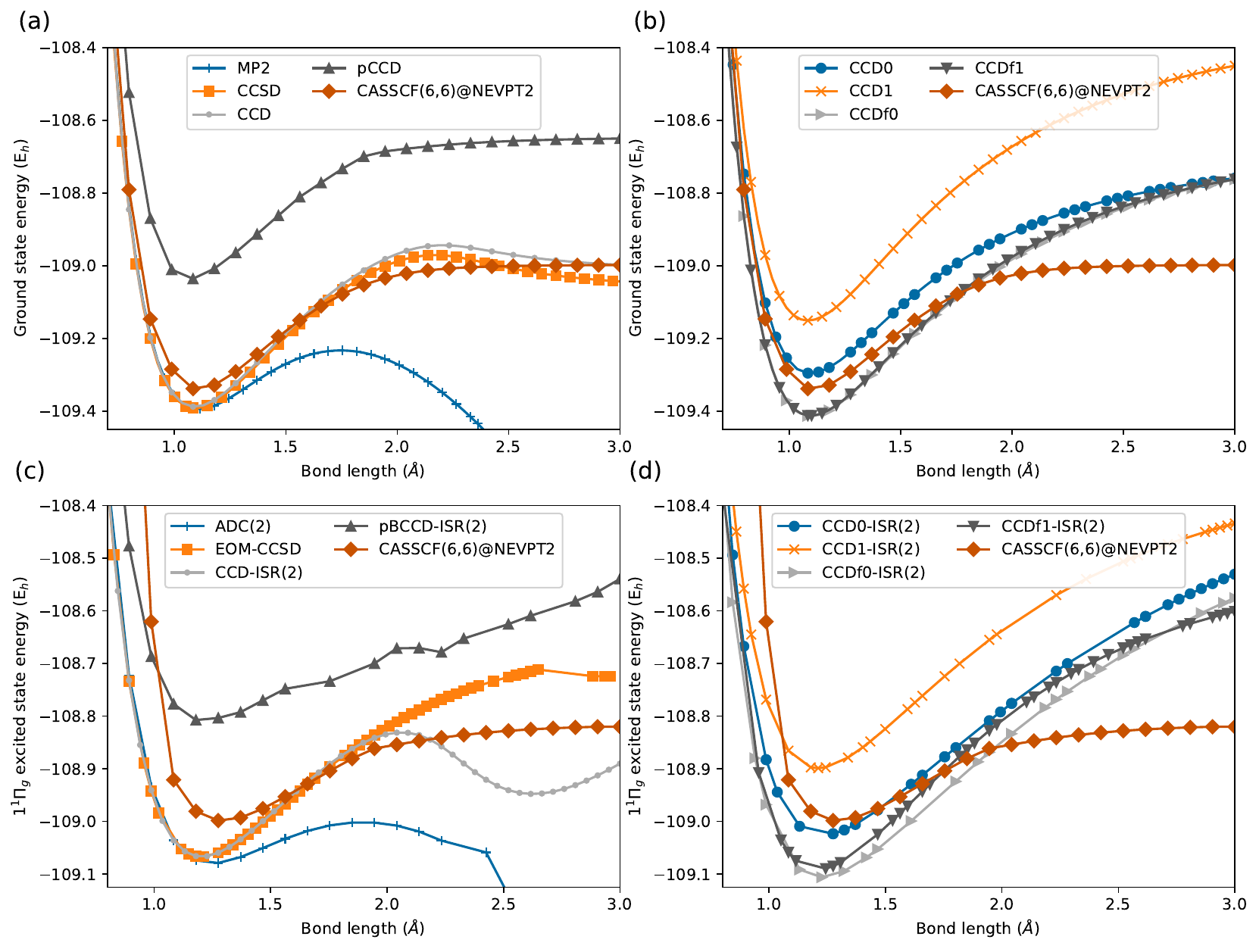}
    \caption{
    Potential energy surfaces
    along the bond-stretching coordinate of
    N$_{\text{2}}$ molecule
    in (a,b) the ground state and (c,d)
    first degenerate singlet state,
    corresponding to a $\pi\rightarrow\pi^\ast$ transition.
    \textcolor{red}{At $R_{\text{NN}} = 0.9$~\AA\ and $1.0$~\AA, the CASSCF@NEVPT2 reference became unstable,
    so these two points are approximated by CASSCF alone without NEVPT2 corrections.}
 }
    \label{fig:N2}
\end{figure*}

Moving towards a particularly challenging
physical system, we now analyze
the performance of various excited state
approaches in the dissociation of N$_{\text{2}}$.
The results in Figures~\ref{fig:N2}a and \ref{fig:N2}b show
the ground state potential
energy surface along the N$\equiv$N bond
stretch coordinate \textcolor{red}{ for which the restricted HF solution obeys N$_{\text{2}}$ spatial symmetry.\cite{RisPerBar16}}
\textcolor{red}{Throughout this section, we take the benchmark
to be CASSCF@NEVPT2 with 6 active electrons and 6 active orbitals for both the
ground- and excited-state potential energy surfaces.}

Again we find that the ground state results
are predictable, with MP2 diverging
and CCD and CCSD methods featuring
a qualitatively incorrect barrier in the potential
surface around 2~\AA\ before the energy lowers
towards the dissociation limit.
Despite being much too high in energy, pCCD even with canonical orbitals 
performs
remarkably well in capturing the qualitative
shape of the potential surface relative to
\textcolor{red}{CASSCF@NEVPT2}.
Singlet-paired CCD0 and triplet-paired CCD1 smoothly dissociate
N$_{\text{2}}$ without any artificial barriers,
but the correlation energy at the equilibrium geometry is
underestimated due to a lack of dynamical
correlation.
The equilibrium energy
is captured more accurately by CCDf0 and CCDf1,
as the perturbative inclusion of the missing
amplitudes accounts for the missing dynamic correlation in CCD0 or CCD1, without reintroducing the
artificial barrier near 2~\AA. In fact, CCDf0 and CCDf1 somewhat over-correlate at the equilibrium geometry in comparison with CCD or CCSD. \textcolor{red}{The equilibrium bond lengths and vibrational frequencies predicted by all of the CC methods except for pCCD and CCD1 are all fairly close to those of CCSD (Table~\ref{tab:spec_param}), indicating that the better-performing addition-by-subtraction CC methods retain the correct shape of the ground state potential energy surface near the equilibrium geometry.}
In agreement with previous work, CCDf0 and CCDf1 perform similarly in the ground state for N$_2$ dissociation.\cite{GomHenScu16}
While they do not diverge,
all addition-by-subtraction CCD methods
overestimate the energy at dissociation. \textcolor{red}{At 12~\AA\ the rough limits achieved
are -108.68 eH, -108.65 eH, and -108.63 eH,
for CCD0, CCDf1, and CCDf0,
respectively.}

The potential surface corresponding to the
$1^1\Pi_g$ excited state
of N$_{\text{2}}$ is shown in Figures~\ref{fig:N2}c and \ref{fig:N2}d.
This is a bound excited state that corresponds
to a $\pi\rightarrow\pi^\ast$ transition, an excitation that reduces
the net bond order but is not sufficient to make the state dissociative.
Once again we find that ADC(2) cannot adequately describe
the excited state potential
surface due to the divergence of the MP2
reference state, rendering ADC(2) useless for
this problem outside the Franck-Condon region.

Despite the deficiencies of CCSD in the reference state,
EOM-CCSD appears to mostly smoothly dissociate
N$_{\text{2}}$ in the excited state,
suggesting that the EOM procedure may
iron out wrinkles in the
excited state surface.
Unlike EOM-CCSD, the
artificial barrier in the
CCD ground state is amplified in
the CCD-ISR(2) excited state,
which appears to be consistent
with previous studies of the approximate
CC-ADC(2) approach.\cite{HodDemReh19}

With canonical HF orbitals, pCCD-ISR(2) predicted the N$_2$ excited state to be basically unbound, as shown in Fig.~\ref{fig:N2_pccd}.
To the contrary, we found that EOM-pCCD+S (even without optimized orbitals)
predicts a bound excited state. 
Again suspecting the importance of $\hat{T}_1$ for pCCD-based excited states, we calculated pBCCD-ISR(2) and pCCSD-ISR(2) excited states, \textcolor{red}{finding
that only the former provided a definitively bound excited state.
This supports our hypothesis that the ground\slash singly-excited state coupling,
which is not properly incorporated in CC-ISR(2) approaches that use canonical orbitals
but is formally disposed of in the Brueckner basis,
is likely essential in modeling the excited state of N$_2$ with pCCD-based methods.}
We note that pCCD provides a \textcolor{red}{less desirable} scaffold
upon which to build a proper excited-state theory, as
unlike all the other CC/CC-ISR(2) methods explored in this work,
pCCD (and hence also pCCD-ISR(2) and EOM-pCCD+S) is not size-consistent
without orbital optimization.\cite{Bog16, Bog19, Bog21} 

Similarly to their ground states,
the ISR(2) approaches based on
CCD0, CCD1, CCDf0, and CCDf1 references
yield smooth excited state surfaces
with no artificial barriers.
\textcolor{red}{Like in the ground state, CCDf0-ISR(2) and CCDf1-ISR(2) achieved
a dissociation limit (-108.59 eH and -108.60 eH, respectively) slightly above that of
CCD0-ISR(2) (-108.63 eH) at 12~\r{A}.}
At the equilibrium geometry, CCDf0-ISR(2) and CCDf1-ISR(2) over-correlate the excited state, as they did for the ground state, while CCDf1-ISR(2) over-correlates to a lesser extent.
In Figures~\ref{fig:N2_t1_ccd0} and \ref{fig:N2_t1_ccdf1}, we examine the importance of the inclusion of $\hat{T}_1$ in the CC wavefunction for ISR(2) excited states based on CCD0 and CCDf1. At equilibrium, all of the approaches provide ground and excited states of similar quality. Like in the Hubbard case, CC with Brueckner orbitals leads to a higher-energy excited state than CC with single excitations.

\textcolor{red}{We further quantify all of our
observations by plotting the non-parallelity error (NPE) 
for the ground and excited states
in Fig.~\ref{fig:N2_NPE} with respect to
the CASSCF@NEVPT2 reference curves.
Apart from supporting what we have already stated,
the NPE plots reveal that, despite a distinct lack of
dynamical correlation that results in a large offset
from the reference energy,
pCCD with Brueckner orbitals appears to
capture the overall curvature of the N$_2$ potential
surfaces quite well.
Nonetheless, without the specific choice of
Brueckner orbitals, pCCD fails dramatically in the
excited state.
}

Overall, the approaches with
smooth ground \emph{and} $^1\Pi_g$ potential
energy surfaces without divergence or anomalous barriers are CCD0\slash CCD0-ISR(2),
CCD1\slash CCD1-ISR(2), CCDf0\slash CCDf0-ISR(2), and CCDf1\slash CCDf1-ISR(2); these methods with the inclusion of $\hat{T}_1$ in some fashion; and pBCCD/pBCCD-ISR(2). Our results suggest that improving the qualitative
nature of the reference wave function
can improve excited state descriptions
within an ISR framework.

\begin{figure*}
    \centering
    \fig{0.8}{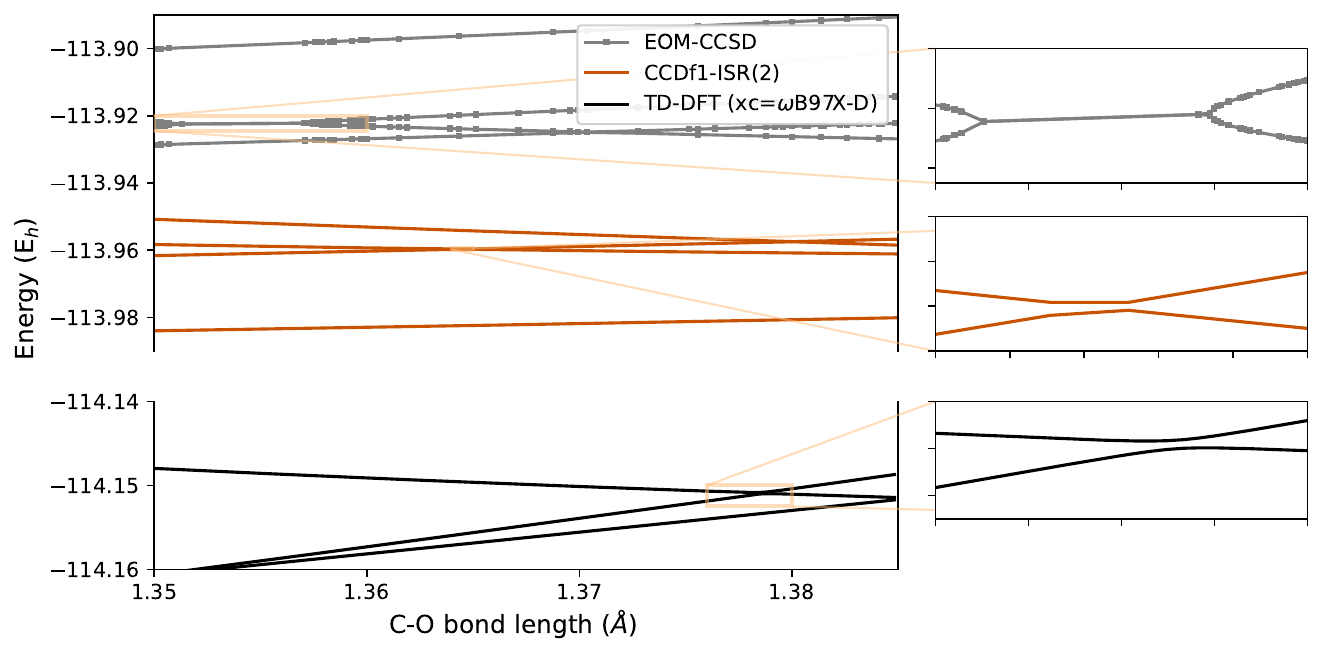}
    \caption{
    Excited state energies for formaldehyde as a function of C$=$O bond length. We used the $\alpha_{\text{OCH}}=118^\circ$,
    $R_{\text{CH}}=111.915$~pm geometry specified in Ref.~\citen{KohTaj07}. The S5-S6 (2 $^1\text{A}_1$-3 $^1\text{A}_1$) avoided crossing predicted by TD-DFT and CCDf1-ISR(2) and incorrect degeneracy region predicted by EOM-CCSD are shown in the insets.
    }\label{fig:CI}
\end{figure*}

\subsection{\textcolor{red}{Formaldehyde PES topology}}
Next, we emphasize some of the more favorable attributes of
ISR methods. Namely, the Hermitian
ISR framework has the capacity to circumvent
several major problems with the EOM-CC
approach. While we will not
focus on this here, one consequence
of the non-Hermitian framework of EOM-CC 
is that the predicted intensities
are not size-intensive.\cite{StaBar93, KocKobDeM94}
Thus,
intensities of local excitations
on a chromophore can be perturbed by the
presence of an atom that is infinitely
far away (yet present) in the calculation.
\textcolor{red}{Notably, linear response CC is an alternative to EOM-CC that, while still non-Hermitian, provides size-intensive intensities,
but being that EOM-CC is more commonly used, we focus on EOM-CC models here.\cite{AhmBogTec24}}

Another problem with the non-Hermitian
formulation of EOM-CC is that conical intersections/avoided crossings between two excited states
of the same symmetry are not correctly
described. The non-Hermitian
EOM-CC Hamiltonian
imparts a lack of orthogonality between the excited states and can lead to
complex solutions to the EOM problem. On the other hand, 
the CC-ISR(2) approach
imposes orthogonality between the excited
states, and the Hamiltonian is
Hermitian in the correlated excited state
basis.
In this sense, CC-ISR(2) methods should
correctly describe the topology
of conical intersections/avoided crossings between excited states
of the same symmetry
and predict size-intensive oscillator
strengths.
\textcolor{red}{Note that
the conical intersection problem between excited states
is distinct from correctly predicting S0\slash S1 conical intersection topology
between ground and excited states.
At the moment, no ISR(2)-based method (apart from the aforementioned
spin-flip variants) can correctly
predict S0\slash S1 conical intersection topology due to
an imbalance in the treatment of ground and excited states,\cite{TunLefWol15}
a problem shared with TD-DFT.}

We assess the ability of CC-ISR(2) to predict potential energy surface
topology \textcolor{red}{of two isosymmetric excited states} by examining
the notorious case of C$=$O bond
stretching in formaldehyde.\cite{KohTaj07}
Although similarity-constrained EOM-CCSD
can predict an avoided
crossing,\cite{KjoKoc17,KjoKoc19}
our standard EOM-CCSD results in Fig.~\ref{fig:CI}
exhibit the expected incorrect topology.
Because TD-DFT correctly captures the topology of
isosymmetric conical intersections between
two excited states, we have included
it as a qualitative metric.
Our TD-DFT calculations reveal that the 2~$^1\text{A}_1$-3~$^1\text{A}_1$
surfaces do indeed undergo an avoided crossing, which is qualitatively
consistent with our CCDf1-ISR(2)
result.
Considering that conical intersection
topology can have a qualitative impact
on photodynamics, this is a clear
advantage of CCDf1-ISR(2) over EOM-CC
approaches.\cite{ZhaHer21}

\subsection{\textcolor{red}{Performance on Quest\#1 Database}}
\begin{figure}
    \centering
    \fig{0.5}{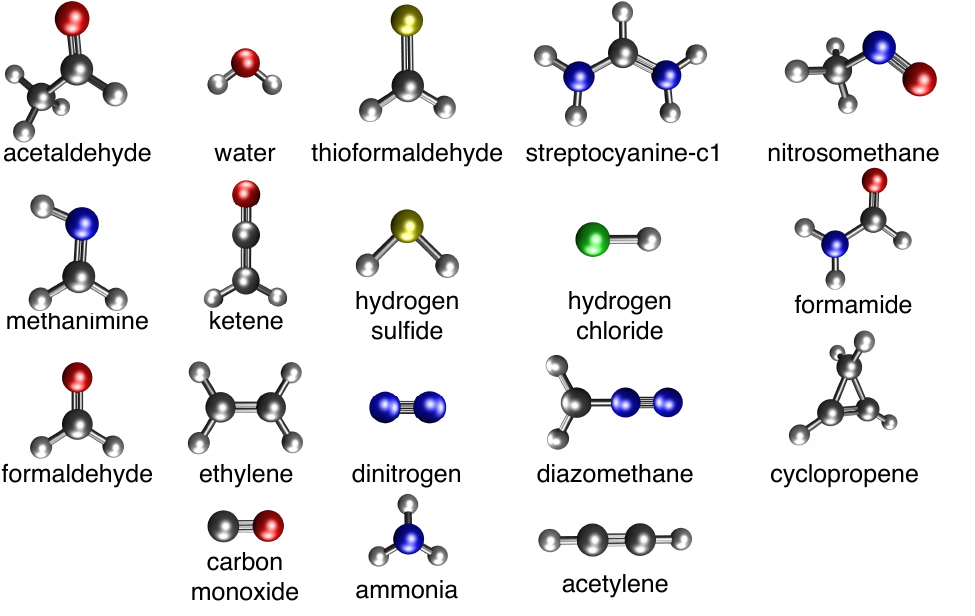}
    \caption{
    The 18 molecules in the Quest \#1 database.\cite{LooSceBlo18} We performed calculations on the 17 closed-shell molecules in this set, which excludes one open-shell molecule (streptocyanine-c1).
    }\label{fig:Quest_molecules}
\end{figure}

\begin{figure}
    \centering
    \fig{1.0}{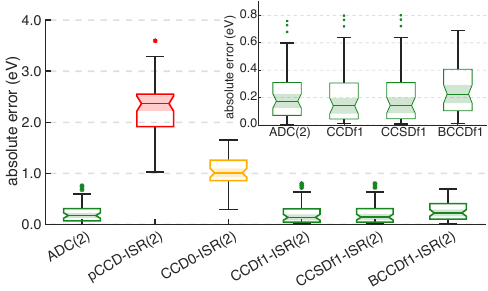}
    \caption{
    Error statistics (in eV) of 52 singlet excitation energies for the 17 closed-shell molecules in the Quest \#1 database.\cite{LooSceBlo18} Excitation energies
    are compared with the aug-cc-pVTZ basis Quest \#1 theoretical best estimates. 
    Upper and lower delimiters indicate the maximum and minimum error,
    while the upper and lower bounds of each box indicate
    the upper and lower quartiles, respectively. Median absolute errors
    are indicated by the central horizontal line and outliers by asterisks.
    Overlapping notches (highlighted by shaded regions) indicate a statistical
    similarity between distributions up to a 95\% confidence interval.
    The ADC(2) data was taken from the Quest \#1 database.
    The inset shows the error distributions for
    ADC(2) and the CC-ISR(2) approaches with lowest
    error in more detail.
    }\label{fig:Quest}
\end{figure}

Thus far we have expounded upon the
capacity for CCDf1-ISR(2) to correctly
describe
potential energy surface topology and
topography with little mention of
the quality of CCDf1-ISR(2) absolute excitation energies.
Our focus on potential surfaces was motivated by the
fact that non-adiabatic photodynamics
simulations depend more strongly
on the shapes of the
potential surfaces ({\em i.e.} derivatives, forces) than they do on
the absolute energies, 
but accurate energy differences at the equilibrium
geometry are also necessary.
To this end, we benchmarked
our suite of CC-ISR(2) approaches
on the QUEST~\#1 organic molecules dataset
of Loos {\em et al.}\cite{LooSceBlo18}, pictured in Fig.~\ref{fig:Quest_molecules}.
The results in Fig.~\ref{fig:Quest}
emphasize the importance of proper
inclusion of dynamical correlation effects
to the computed excitation energy, as
our CC-ISR approaches exhibit mean
absolute errors (MAE)
that decrease monotonically with
greater inclusion of dynamical correlation.
Specifically, we find MAE of
2.3~eV, 1.0~eV, and 0.21~eV
for pCCD-ISR(2), CCD0-ISR(2),
and CCDf1-ISR(2), respectively. 
\textcolor{red}{We note that pCCD-ISR(2) and EOM-pCCD+S even with canonical orbitals can incorrectly break degeneracies (Table~\ref{tab:degen_pccd}). We excluded five pCCD-ISR(2) excitation energies (for CO and N$_2$) because the predicted degeneracy pattern was incorrect.}
Our BCCDf1-ISR(2) approach (MAE 0.26~eV)
is comparable in accuracy to ADC(2) (MAE 0.23~eV),
which is consistent with the results
reported in Table~III of Ref.~\citen{HodRehDre20}
for CCD-ISR(2). 

We note that CCDf0-ISR(2) gives a slightly larger MAE of 0.28 eV. This is likely because the triplet amplitudes represent a smaller
fraction of the total correlation energy beyond HF, so solving for those first,
then freezing them and computing the infinite-order solution
to Eq.~\ref{eq:xCC} ensures that the largest component of the correlation
energy is iteratively optimized with at least approximate
external amplitudes provided by the triplet contribution.
Solution of the CCDf0 equations imparts more error by approximating
the larger (singlet) contribution in the absence of other amplitudes
while the infinite-order external CC equation for the remaining triplet
component only offers small corrections.

Including $\hat{T}_1$ through CCSDf1-ISR(2) did not affect the quality of the computed Quest \#1 excitation energies, giving a MAE of 0.21 eV.
While BCCDf1-ISR(2) 
gave a slightly increased MAE of 0.26 eV, the error margins
are statistically the same as ADC(2) and other approaches,
as indicated by the overlapping shaded regions
in the inset of Figure~\ref{fig:Quest}.
Interestingly, using the slightly more justified Brueckner determinant
and corresponding 
$t_2$ amplitudes as a starting
point for ISR(2) results in the elimination of outliers in the
statistics, meaning the BCCDf1-ISR(2) provides a more
even-handed description across all varieties of excited state
represented in Quest\#1.
Overall, our Quest\#1 benchmarking results suggest that we
can use BCCDf1-ISR(2) to attain the same level of accuracy
in excitation energies as ADC(2)
and full CCD-ISR(2) but with
a better overall treatment of
static correlation, manifesting in
potential energy surface topographies that
are significantly more accurate than those provided by either of the latter approaches.

\subsection{\textcolor{red}{Double excitations}}
\textcolor{red}{Quantitatively accurate models of excitations with substantial double excitation character generally require wavefunction methods containing at least triple substitutions, such as a multireference CAS-based method or ADC(3).
While our CC-ISR(2) approaches do not contain triple excitations, they do model
double excitations to zeroth order in the energy,
permitting at least some coverage of the doubles manifold.
To understand the extent to which CC-ISR(2) approaches can account for double excitations,
we assessed the performance of CCSDf1-ISR(2) for several linear polyenes.
Such trans-polyenes host $^1$A$_g$ ($\pi\rightarrow\pi^\ast$) states that are known to have substantial double-excitation character.
Table~\ref{tab:doubles} shows excitation energies for ethene, butadiene, hexatriene, and octatetraene,
as calculated with the Def2-TZVPP basis set.
As expected, CCSDf1-ISR(2) performs slightly better than ADC(2) (by about 0.1 eV) even for double excitations and seems to be on par with EOM-CCSD for
the $^1 A_g$ states while outperforming EOM-CCSD for the $^1B_{1u}$ states.
These results suggest that improving the ground-state reference can impart improvements
to the predicted excitation energies of transitions
that have pronounced double-excitation character.
} 
\begin{table}[h]
\caption{Excitation energies (eV) 
for four alkenes}
\setlength{\tabcolsep}{1pt}
\label{tab:doubles}
\begin{tabular}{ll....}
\hline\hline
\mc{1}{c}{system}   & \mc{1}{c}{state} & \mc{1}{c}{EOM-CCSD}	& \mc{1}{c}{ADC(2)}	& \mc{1}{c}{CCSDf1-ISR(2)} & \mc{1}{c}{TBE$^a$} \\ 
   \hline
 C$_2$H$_{4}$ & 1 $^1$B$_{1u}$ & 8.31 &8.21	& 8.11	&7.80\\  
 \hline
 \multirow{2}{*}{C$_4$H$_{6}$} & 1 $^1$B$_{1u}$ & 6.55	& 6.28	& 6.17	& 6.18\\
 
    & 2 $^1$A$_g$ & 7.45 & 	7.59 & 	7.48	& 6.55\\ 
    \hline
\multirow{2}{*}{C$_6$H$_{8}$} & 1 $^1$B$_{1u}$ & 5.55	& 5.22	& 5.09	& 5.10\\  
 
   & 2 $^1$A$_g$ & 6.66 &	6.65	& 6.54	& 5.09\\
   \hline
\multirow{2}{*}{C$_8$H$_{10}$} & 2 $^1$A$_g$ & 6.02	& 5.87 &	5.74	& 4.47\\  
   & 1 $^1$B$_{1u}$ & 4.91	& 4.53	& 4.40	& 4.66\\
   \hline
   MAE & & 0.80 & 0.64 & 0.53 & \\
   \hline\hline
\mc{6}{l}{\fns $^a$Theoretical best estimates from Ref.~\citen{SchSilSau08}}
\end{tabular}
\end{table}


\section{Conclusions and Outlook}
We have introduced addition-by-subtraction coupled
cluster (CC) with double substitutions (CCD)
approaches as an {\em ansatz}
for the intermediate state representation (ISR)
approach for calculating excited states
in order to account for nondynamical correlation
with polynomial-scaling, single-determinant
reference wave functions.
Among those tested were
pair CCD (pCCD), singlet-paired
CCD (CCD0), triplet-paired CCD (CCD1), 
and CCD with frozen singlet/triplet-paired amplitudes
(CCDf0/CCDf1), as well as these methods with CC-singles excitations or Brueckner instead of canonical HF orbitals.

\textcolor{red}{Of course, there will always be limitations
in the single-reference description of static correlation,\cite{ChaKalGau04, BogTecBar13}
but }
all \textcolor{red}{CC-ISR(2)} approaches except for pCCD-ISR(2) with canonical orbitals
were capable of correctly predicting the topography of excited state potential energy surfaces
along bond dissociation coordinates
with qualitative accuracy.
For quantitative accuracy of excitation energies
and qualitatively accurate potential surfaces,
we recommend BCCDf1-ISR(2), as it captures a greater fraction of dynamical correlation
and predicts excitation energies with statistical
accuracy on par with that
of second-order algebraic diagrammatic construction
[ADC(2)] but without the dramatic failures often
exhibited by ADC(2) when the underlying
second-order M{\o}ller-Plesset (MP2)
reference energy diverges.
Importantly, 
BCCDf1-ISR(2) is somewhat more formally sound
than its CCDf1-ISR(2) \textcolor{red}{or CCSDf1-ISR(2)} counterparts, as it
rigorously eliminates singles\slash reference-state
coupling that should otherwise appear at each
order in the ISR equations.

The Hermitian construction of the ISR excited states
also allows CCDf1-ISR(2) to describe avoided crossings
and conical intersections correctly where
equation-of-motion (EOM) CC methods may fail,
albeit at a slight sacrifice of accuracy in the
excitation energies.
While CCDf1-ISR(2) itself may not enter widespread use,
we believe that future adaptations
of the approach will be useful for modeling
photodynamics due to their capacity to correctly
describe potential energy surface topology and topography.

Some of the most important contributions of this work
are conceptual advancements that were made in
our first steps towards finding optimal \textcolor{red}{single-}reference
excited state theories.
Firstly, that some -- but not all -- reference wave functions
that provide qualitatively good descriptions of
statically-correlated systems in their electronic
ground state can also improve the quality of
the predicted excited states.
Especially important is the remarkable failure of pCCD-ISR(2) in the dissociation of N$_{\text{2}}$,
where the ground state pCCD topography was perhaps among
the best of the methods that we tested.
\textcolor{red}{While this may not necessarily be the case for EOM- or linear response-based approaches,} within the ISR formulation, pCCD reference wave
functions with canonical HF MOs 
do not even qualitatively
capture static correlation in excited states.
\textcolor{red}{Although these failures can be somewhat alleviated
by choosing the Brueckner orbital basis, our results}
emphasize
that caution should be exercised when calculating excited states
with methods that are not invariant to unitary transformations of
the orbital basis within the occupied or virtual spaces.

We also demonstrated the usefulness of the ISR procedure
in
improving the description of an avoided
crossing. This motivates our further study into a more
formal derivation of an ISR framework from a variety of CC
reference
wave functions employing a Hermitian framework. 
A more rigorous approach than CC-ISR(2) should treat the ground and excited state wave function at the same level of approximation. (Preliminary testing suggests that this may be necessary for the quantitative treatment of absolute excitation energies in transition metal-containing molecules.) 
Additionally, we are actively investigating related CC approaches
that may avoid the $\mathcal{O}(N^6)$ scaling of
CCDf1-ISR(2), which is bottlenecked by the solution
of the CCDf1 amplitudes. With more concrete formalism and lower computational
cost, methods similar to CCDf1-ISR(2) may offer efficient yet robust
alternatives to complete active-space approaches to calculating
photodynamics of
strongly correlated excited states that typify
common processes such as photolysis.
Most importantly, it
would seem that the general concept of
optimal \textcolor{red}{single}-reference
theories for excited states
(the idea that improvements in the initial reference 
wave function can beget improvements to the predicted excited states)
is well-founded
and merits further research.
\section*{Acknowledgements}
We thank Ethan Vo and Nastasia Mauger for assistance with PySCF installation issues, Md. Rafi Ul Azam for help with processing QChem output, and Samragni Bannerjee for providing the EE-ADC(2) PySCF code by Terrence Stahl and Alexander Sokolov.
This research was supported in part by the University of Pittsburgh
and the University of Pittsburgh Center for Research Computing, RRID:SCR\_022735, through the resources provided.
Specifically, this work used the H2P cluster, which is supported by NSF award number OAC-2117681.

\newpage

\section*{References}

\bibliography{allbib}
\bibliographystyle{acs_w_titles}

\clearpage
For Table of Contents Only
\\
\includegraphics[]{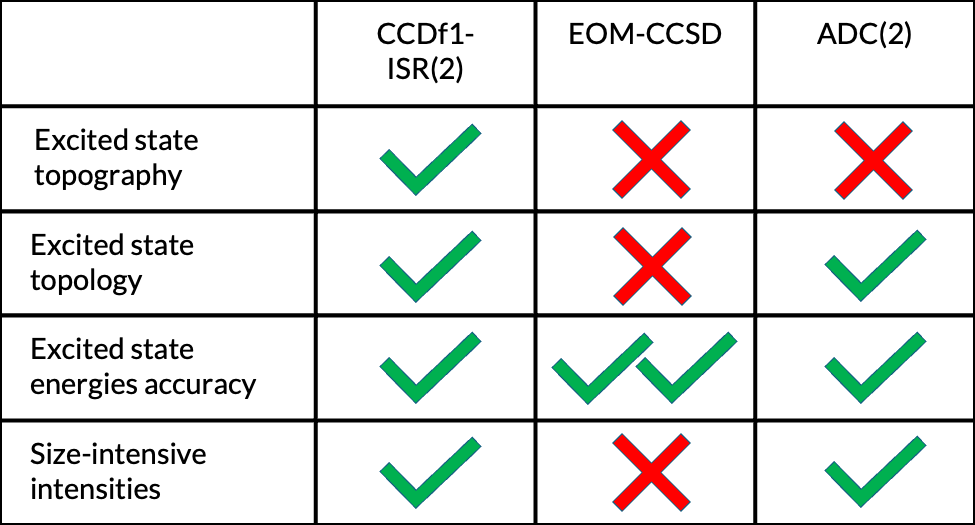}
\end{document}


\title{
Supporting Information for 
\\
Optimal-Reference Excited State Methods:
Static Correlation at Polynomial Cost with
Single-Reference Coupled-Cluster Approaches
}

\author{Sylvia J. Bintrim}

\author{
    Kevin Carter-Fenk
}

\date{\today}

\maketitle
\thispagestyle{plain}

\section{Excited state character in the Hubbard model}
One might ask whether the `kink' in the excitation energy plot near $U/|t|=2$ indicates the presence of a higher-energy singlet excitation with the same orbital character as the excited state curve we started on at low $U$. We show that this is not the case. 
As can be seen in Table~\ref{tab:SI-Hubbard-CI}, the main determinants (and their coefficients' signs) contributing to the S0 ground and
S1 state remain the same as a function of interaction strength from $U/|t|=1$ to $U/|t|=3$. 
The ground state RHF MO coefficients for the HOMO, LUMO, and LUMO+1 are independent of interaction strength. 
Because the ground and excited state characters as well as the MO energetic ordering are independent of $U/|t|$ across the `kink' in the excitation energy around $U/|t|=2$, we can conclude that we are, in fact, tracking the same first-singlet excited state as a function of interaction strength.

There is a notable enhancement of collective character in the ground and excited states even at interaction strengths as low as $U/|t|=3$. So, while we are confident in our
state assignment for the lowest-energy excited state (the ordering of states does not change with interaction strength), we note that the excited states of the Hubbard model provide an exacting test for how well a method
can deal with collective excitation character.
All of the methods that we test in the main text can capture at least {\em some} plasmonic
character, as they explicitly account for double-excitations.

\begin{figure}[h!!]
    \centering
    \fig{0.45}{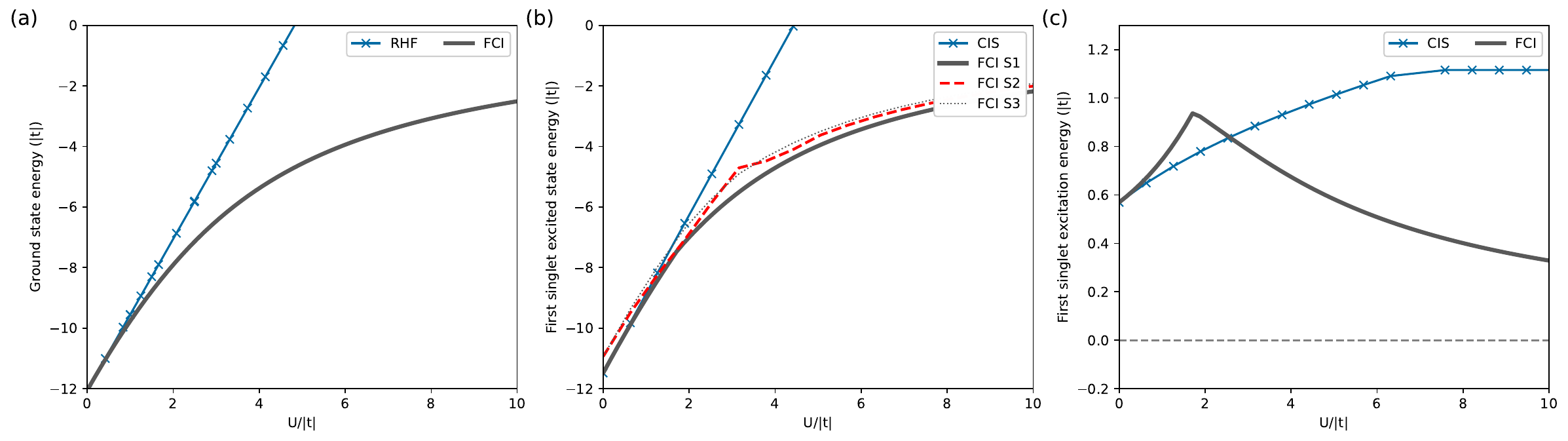}
    \caption{
    (a) Ground state ($E_{\text{S}_0}$),
    (b) singlet excited states,
    and (c) first singlet
    excitation ($E_{\text{S}_1} - E_{\text{S}_0}$) energies
    as a function of interaction
    strength ($U/|t=-1.5|$) for a 10-site,
    half-filled Hubbard model
    with open boundary conditions.
    The FCI result is exact for both
    states and acts as the reference.}
    \label{fig:Hubbard_CIS}
\end{figure}

In Figure~\ref{fig:Hubbard_CIS}, we show that even CIS can predict the kink in the excitation energy plot, though it occurs at much higher $U/|t|$ than predicted by other methods due to the lack of inclusion of doubles-and-higher substitutions in CIS. 
\begin{table}[h]
\caption{Configuration Interaction Coefficients for the Hubbard Model
as a Function of Interaction Strength}
\label{tab:SI-Hubbard-CI}
\begin{minipage}{.45\linewidth}
\begin{tabular}{.c..}
\mc{2}{c}{Ground state}\\
\hline\hline
\mc{1}{c}{$|U/t|$} & \mc{1}{c}{Dominant Configurations} & \mc{1}{c}{$C$} & \mc{1}{c}{$|C^2|$}\\
\hline\\[-1.5ex]
\multirow{2}{*}{1.0} & $ijklm,\bar{i}\bar{j}\bar{k}\bar{l}\bar{m}$ & 0.969 & 0.939\\[5pt]
 & $ijkla,\bar{i}\bar{j}\bar{k}\bar{l}\bar{a}$ (double) & -0.116 & 0.014\\
 \hline\\[-1.5ex]
\multirow{4}{*}{3.0} & $ijklm,\bar{i}\bar{j}\bar{k}\bar{l}\bar{m}$ & 0.754 & 0.568\\[5pt]
 & $ijkla,\bar{i}\bar{j}\bar{k}\bar{l}\bar{a}$ (double) & -0.256 & 0.066\\[5pt]
 & $ijkma,\bar{i}\bar{j}\bar{k}\bar{l}\bar{b}$ & 0.132 & 0.017\\[5pt]
 & $ijklb,\bar{i}\bar{j}\bar{k}\bar{m}\bar{a}$ & 0.132 & 0.017\\
 \hline\\[-1.5ex]
 \multirow{3}{*}{20} & $ijklm,\bar{i}\bar{j}\bar{k}\bar{l}\bar{m}$ & & 0.038\\[5pt]
 & $ijkla,\bar{i}\bar{j}\bar{k}\bar{l}\bar{a}$ (double) & & 0.014\\[5pt]
 & $ijklb,\bar{i}\bar{j}\bar{k}\bar{m}\bar{a}$ & & 0.008\\
\hline\hline
\end{tabular}
\end{minipage}
\begin{minipage}{.45\linewidth}
\begin{tabular}{.c..}
\mc{2}{c}{First singlet excited state}\\
\hline\hline
\mc{1}{c}{$|U/t|$} & \mc{1}{c}{Dominant Configurations} & \mc{1}{c}{$C$} & \mc{1}{c}{$|C^2|$}\\
\hline\\[-1.5ex]
\multirow{2}{*}{1.0} & $ijklm,\bar{i}\bar{j}\bar{k}\bar{l}\bar{a}$ (single) & -0.691 & 0.478 \\[5pt]
 & $ijkla,\bar{i}\bar{j}\bar{k}\bar{l}\bar{m}$ (single) & -0.691  & 0.478\\[5pt]
&
$ijklm,\bar{i}\bar{j}\bar{k}\bar{l}\bar{b}$ (single) & -0.051 & 0.003\\[5pt]
 & $ijklb,\bar{i}\bar{j}\bar{k}\bar{l}\bar{m}$ (single) &  -0.051 & 0.003\\
 \hline\\[-1.5ex]
\multirow{4}{*}{3.0} & $ijklm,\bar{i}\bar{j}\bar{k}\bar{l}\bar{a}$ (single) & -0.603 & 0.364 \\[5pt]
 & $ijkla,\bar{i}\bar{j}\bar{k}\bar{l}\bar{m}$ (single) &  -0.603 & 0.364\\[5pt]
&
$ijklm,\bar{i}\bar{j}\bar{k}\bar{l}\bar{b}$ (single) & -0.105 & 0.011\\[5pt]
 & $ijklb,\bar{i}\bar{j}\bar{k}\bar{l}\bar{m}$ (single) & -0.105 & 0.011\\
 \hline\\[-1.5ex]
 \multirow{1}{*}{20} & $ijkla,\bar{i}\bar{j}\bar{k}\bar{l}\bar{a}$ (double) & -0.123 & 0.015\\[5pt]
\hline\hline
\end{tabular}
\end{minipage}
\end{table}

\newpage
\section{pCCD-based methods for the Hubbard model}
\begin{figure}[h]
    \centering
    \fig{0.6}{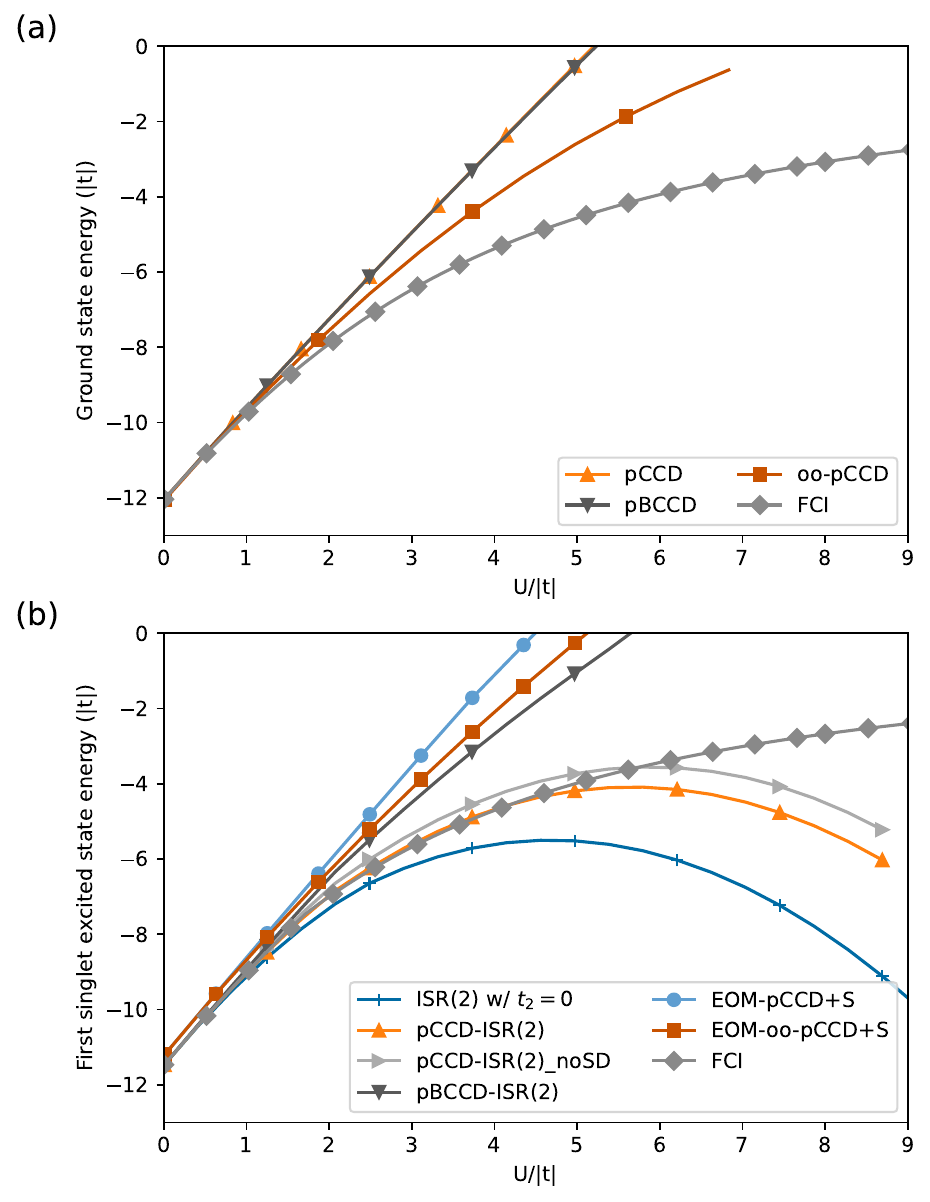}
    \caption{
    (a) Ground state ($E_{\text{S}_0}$) and
    (b) first singlet excited state ($E_{\text{S}_1}$)
    as a function of interaction
    strength ($U/|t=-1.5|$) for a 10-site,
    half-filled Hubbard model
    with open boundary conditions.
    The FCI result is exact for both
    states and acts as the reference. `noSD' denotes that singles-doubles coupling is set to zero. 
    }
    \label{fig:Hubbard_pccd_SD}
\end{figure}

\newpage
\section{EOM-CCSD0 for the Hubbard model}
\begin{figure}[h]
    \centering
    \fig{0.6}{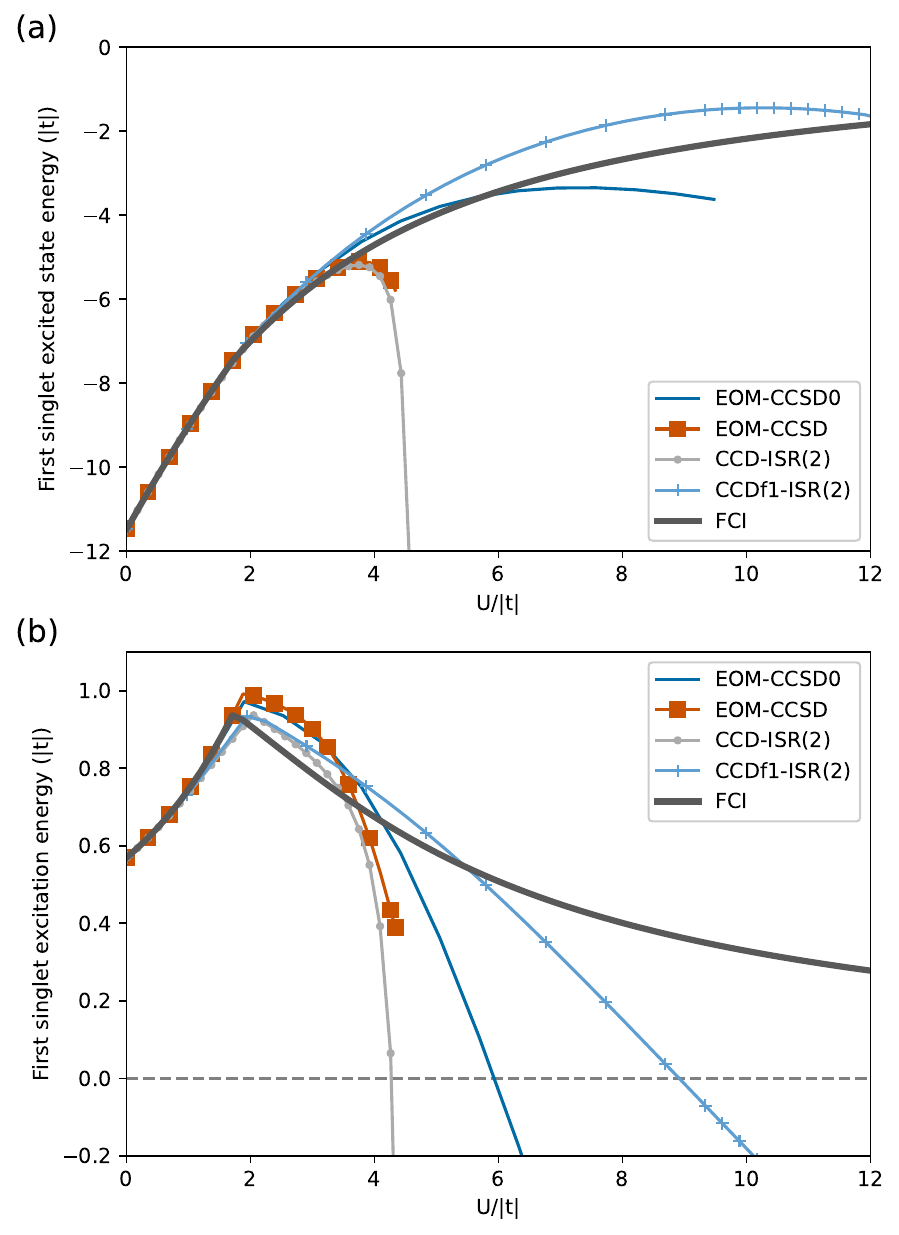}
    \caption{
    (a) First singlet excited state ($E_{\text{S}_1}$) and
    (b) first singlet excitation energy ($E_{\text{S}_1}-E_{\text{S}_0}$)
    as a function of interaction
    strength ($U/|t=-1.5|$) for a 10-site,
    half-filled Hubbard model
    with open boundary conditions.
    The FCI result is exact and acts as the reference.  
    }
    \label{fig:Hubbard_eom_ccsd0}
\end{figure}
\newpage
\section{pCCD-based methods for N$_2$ dissociation}
\begin{figure}[h]
    \centering
    \fig{0.6}{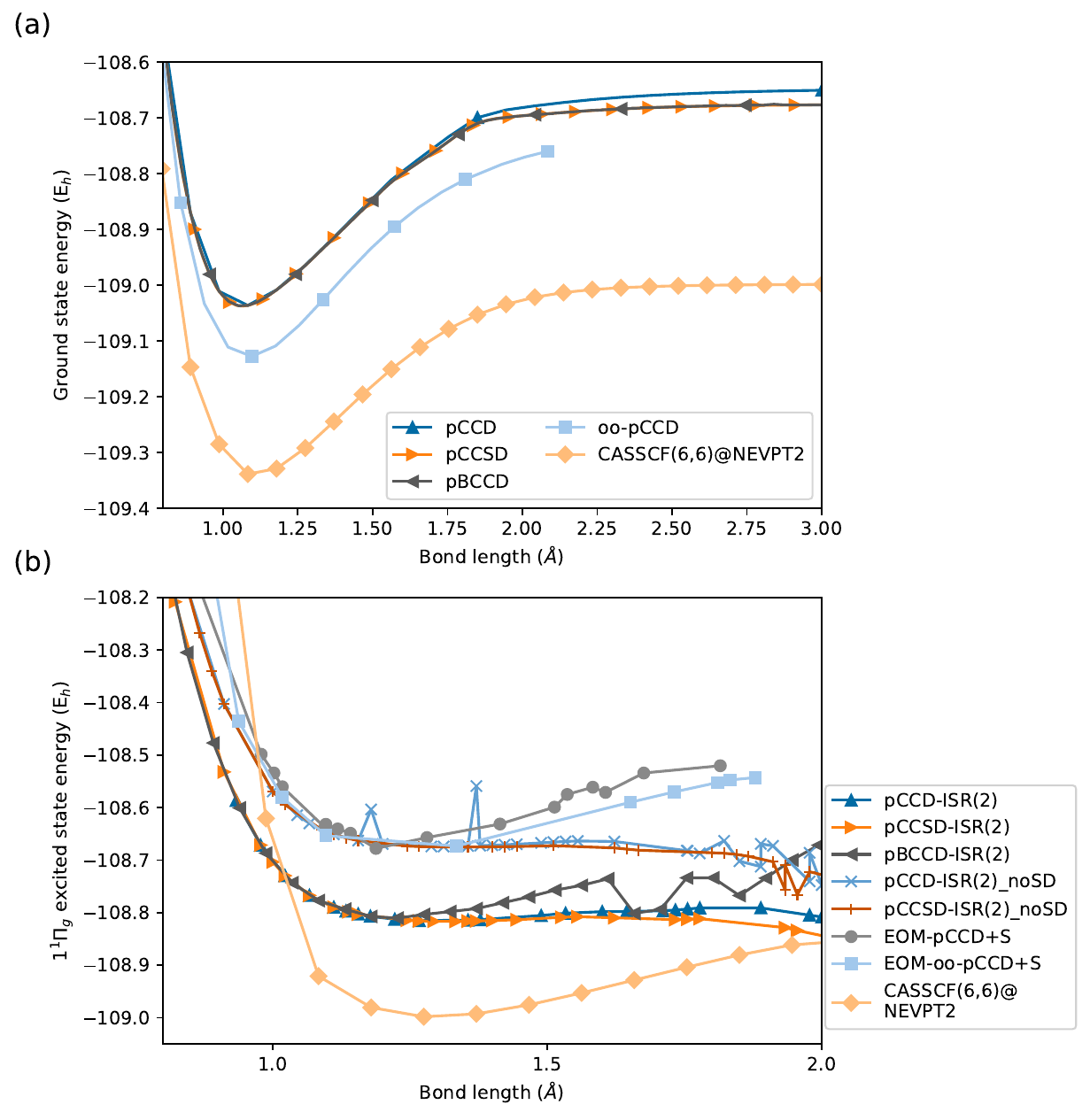}
    \caption{
    Potential energy surfaces
    along the bond-stretching coordinate of
    N$_{\text{2}}$ molecule
    in (a) the ground state and (b)
    first degenerate singlet state,
    corresponding to a $\pi\rightarrow\pi^\ast$ transition.
    `noSD' denotes that singles-doubles coupling is set to zero.
     }
    \label{fig:N2_pccd}
\end{figure}
\newpage
\section{CCD0/CCD0-ISR(2) with $\hat{T}_1$: choices for N$_2$ dissociation}
\begin{figure}[h]
    \centering
    \fig{0.6}{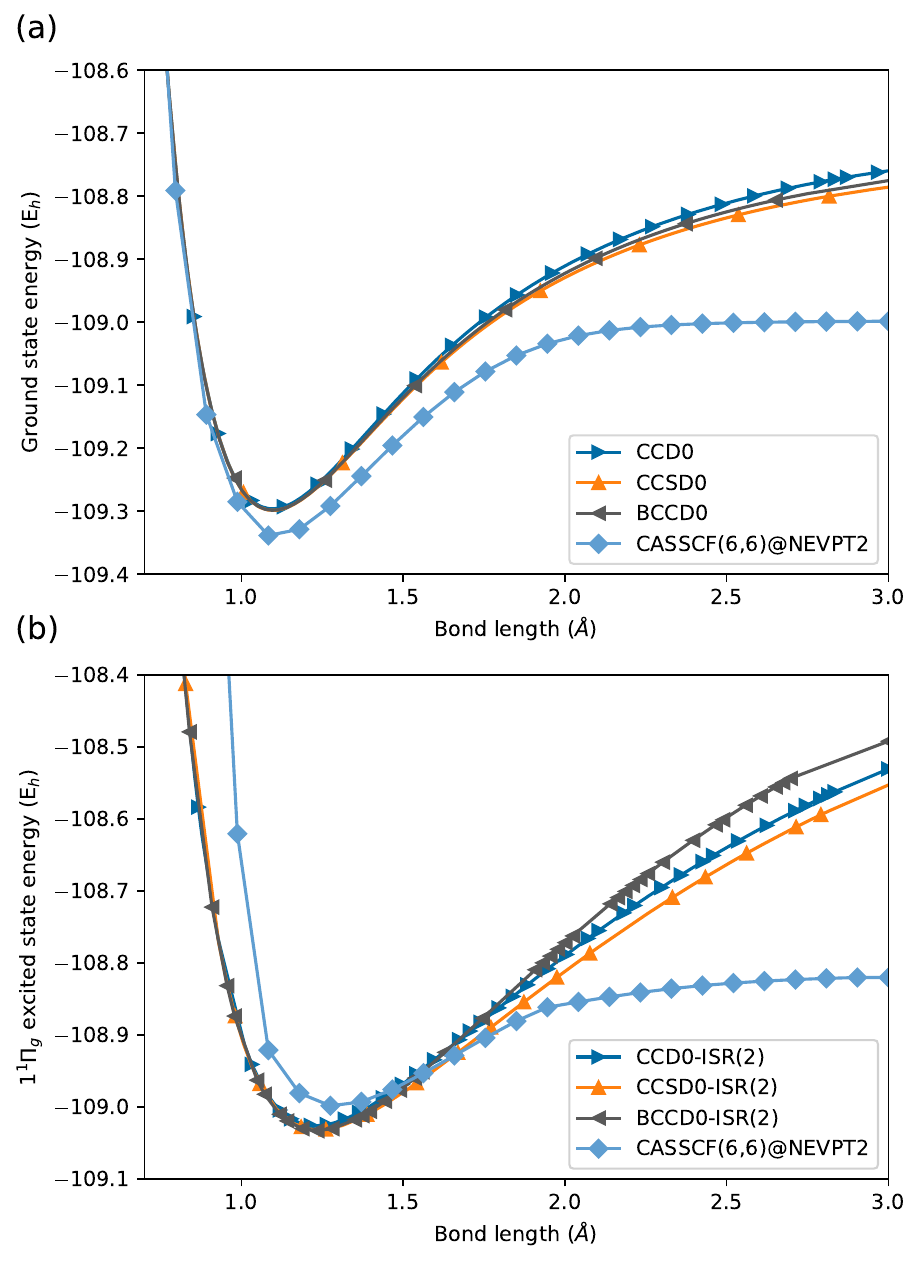}
    \caption{
    Potential energy surfaces
    along the bond-stretching coordinate of
    N$_{\text{2}}$ molecule
    in (a) the ground state and (b)
    first degenerate singlet state,
    corresponding to a $\pi\rightarrow\pi^\ast$ transition.
     }
    \label{fig:N2_t1_ccd0}
\end{figure}
\newpage
\section{CCDf1/CCDf1-ISR(2) with $\hat{T}_1$: choices for N$_2$ dissociation}
\begin{figure}[h]
    \centering
    \fig{0.6}{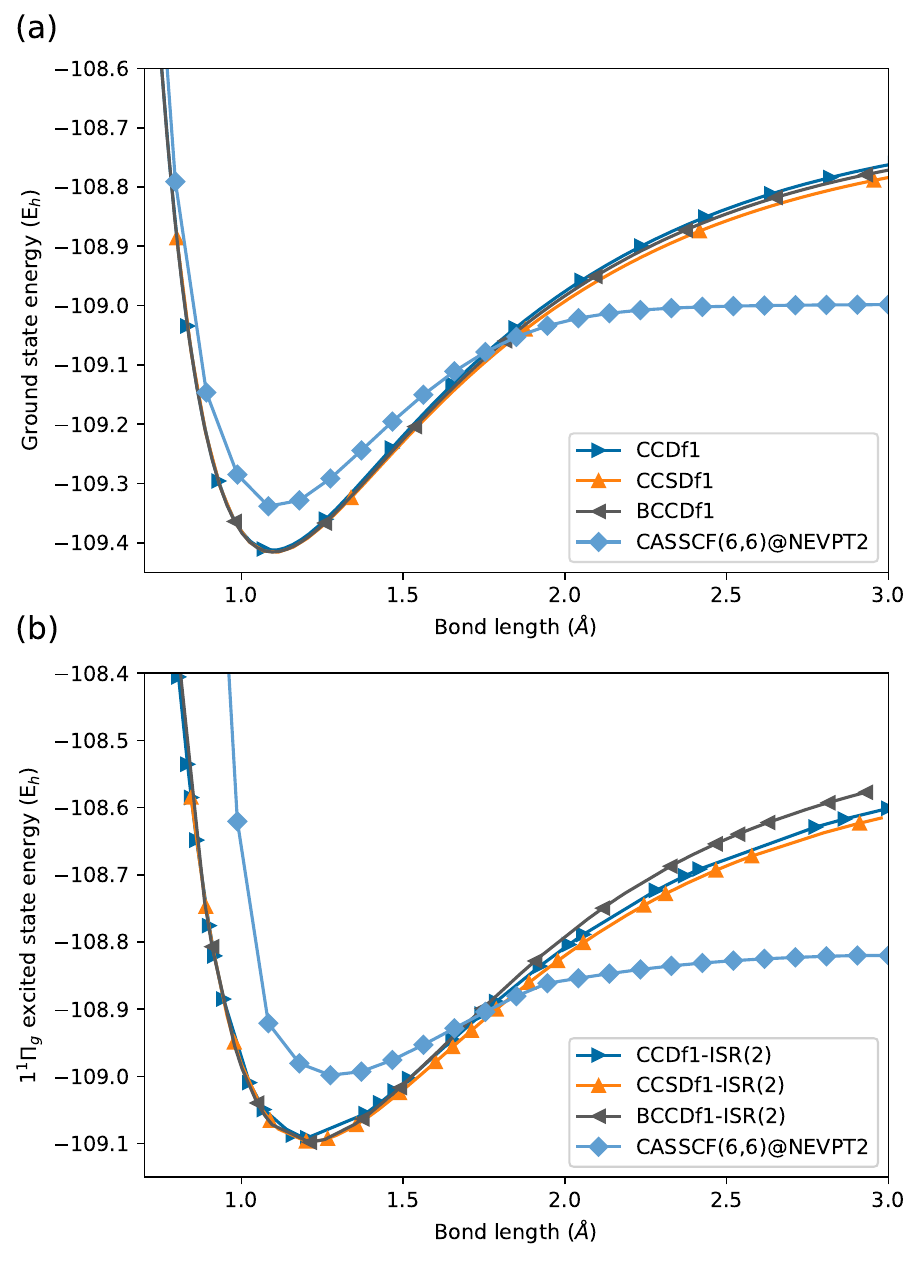}
    \caption{
    Potential energy surfaces
    along the bond-stretching coordinate of
    N$_{\text{2}}$ molecule
    in (a) the ground state and (b)
    first degenerate singlet state,
    corresponding to a $\pi\rightarrow\pi^\ast$ transition.
     }
    \label{fig:N2_t1_ccdf1}
\end{figure}

\newpage
\section{\textcolor{red}{Non-parallelity error plots for N$_2$ dissociation}}
\begin{figure}[h]
    \centering
    \fig{0.5}{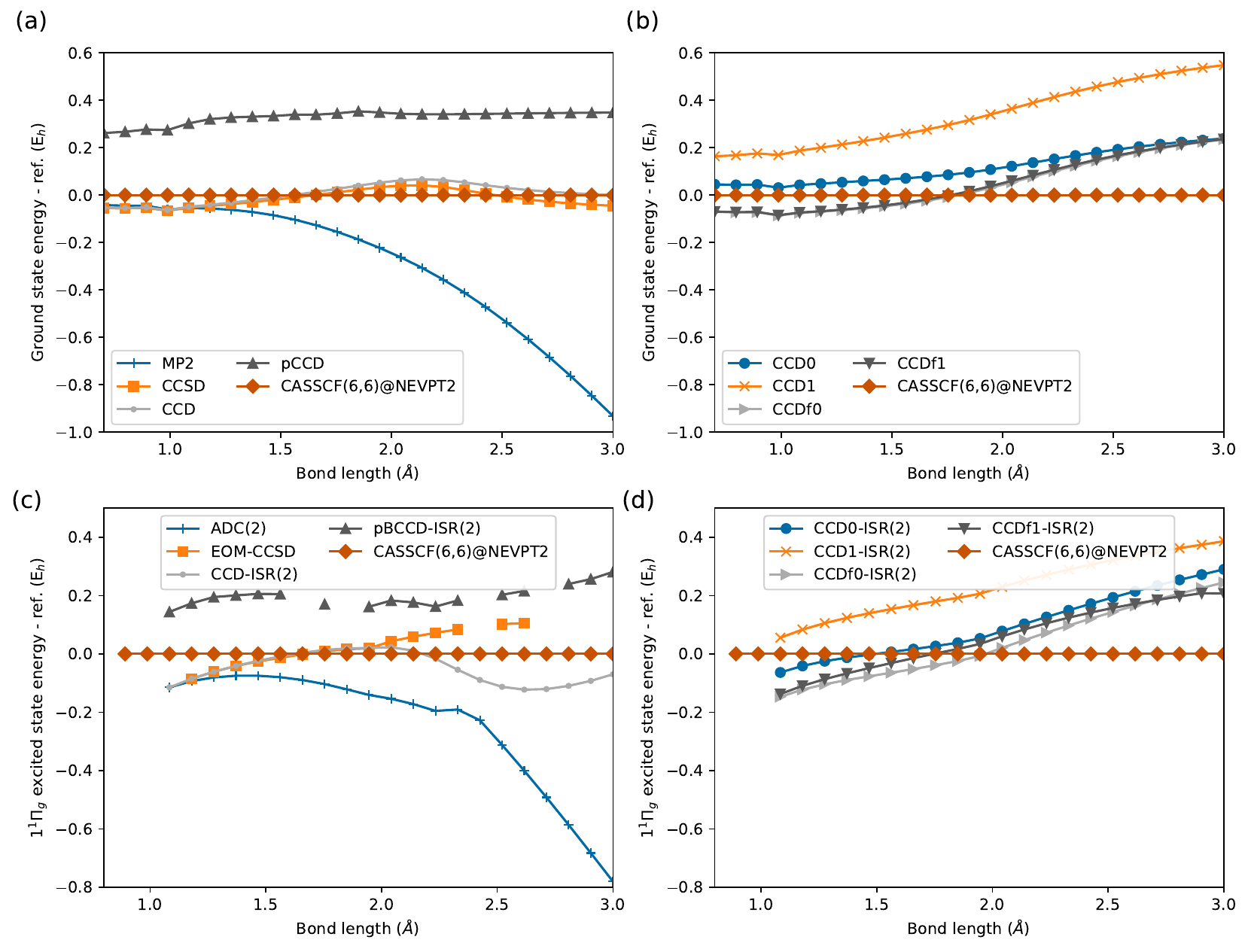}
    \caption{
    Deviation of potential energy surface from the reference potential energy surface (CASSCF(6,6)@NEVPT2)
    along the bond-stretching coordinate of
    N$_{\text{2}}$ molecule
    in (a,b) the ground state and (c,d)
    first degenerate singlet state,
    corresponding to a $\pi\rightarrow\pi^\ast$ transition.
     }
    \label{fig:N2_NPE}
\end{figure}

\newpage
\section{\textcolor{red}{Basis set dependence of N$_2$ dissociation}}
\begin{figure}[h]
    \centering
    \fig{0.5}{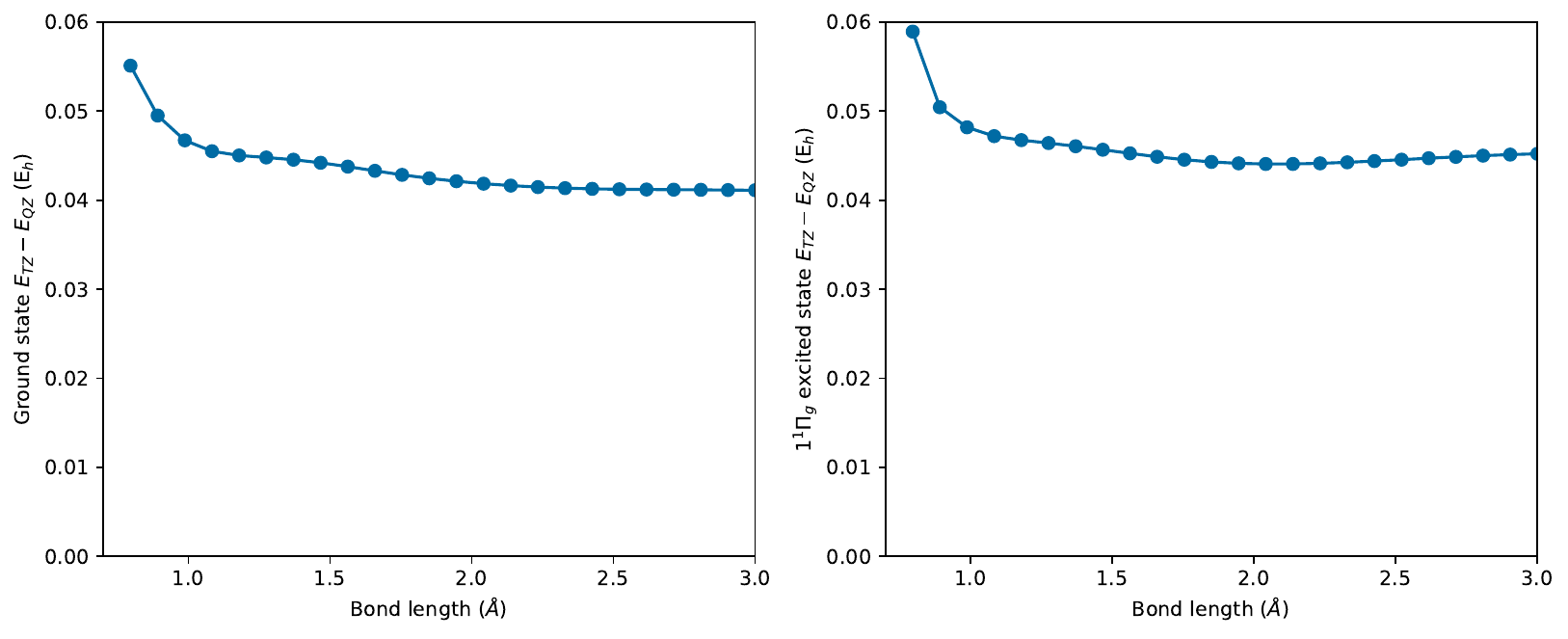}
    \caption{
    Deviation between CCD0/CCD0-ISR(2) potential energy surfaces computed in the aug-cc-pVTZ and aug-cc-pVQZ bases
    along the bond-stretching coordinate of the
    N$_{\text{2}}$ molecule
    in (a) the ground state and (b)
    first degenerate singlet state,
    corresponding to a $\pi\rightarrow\pi^\ast$ transition.
     }
    \label{fig:N2_NPE_basis}
\end{figure}

\newpage
\section{\textcolor{red}{Basis set dependence of alkene excitations energies}}
\begin{table}[h]
\caption{Excitation energies (eV)  
for alkenes in the def2-(SVP/TZVP/QZVP) bases}
\label{tab:doubles}
\begin{tabular}{cc|l|l|l|c}
\hline\hline
\mc{1}{c}{system}   & \mc{1}{c}{state} & \mc{1}{c}{EOM-CCSD}	& \mc{1}{c}{ADC(2)}	& \mc{1}{c}{CCSDf1-ISR(2)} & \mc{1}{c}{TBE$^a$} \\ 
   \hline
 C$_2$H$_{4}$ & 1 $^1$B$_{1u}$ &8.94/8.33/8.15 &8.80/8.23/8.06	& 8.70/8.14/7.97	&7.80\\  
 \hline
 \multirow{2}{*}{C$_4$H$_{6}$} & 1 $^1$B$_{1u}$ & 6.97/6.56/6.44	& 6.66/6.30/6.19	& 6.53/6.19/6.09	& 6.18\\
 
    & 2 $^1$A$_g$ & 7.61/7.46/7.33 & 	8.01/7.61/7.33 & 	7.88/7.50/7.32	& 6.55\\ 
    \hline
\multirow{2}{*}{C$_6$H$_{8}$} & 1 $^1$B$_{1u}$ & 5.89/5.56/	& 5.50/5.23/	& 5.36/5.10/	& 5.10\\  
 
   & 2 $^1$A$_g$ & 6.70/6.66/ &	6.89/6.66/	& 6.76/6.55/	& 5.09\\
   \hline
\multirow{2}{*}{C$_8$H$_{10}$} & 2 $^1$A$_g$ & 6.05/6.02/ & 6.06/5.87/ &			5.92/5.75/	& 4.47\\  
   & 1 $^1$B$_{1u}$ & 5.20/4.92/&			4.77/4.54/ &			4.62/4.40/	& 4.66\\
   \hline
   MAE & & 1.07/0.81/ & 0.98/0.69/ & 0.86/0.61/ & \\
   \hline\hline
   \mc{6}{l}{\fns $^a$Theoretical best estimates (TBE) from Ref.~\citen{SchSilSau08}}
\end{tabular}
\end{table}
Mean absolute errors (MAE) were computed with respect to the TBE. 
\begin{figure}[h!!]
    \centering
    \fig{0.45}{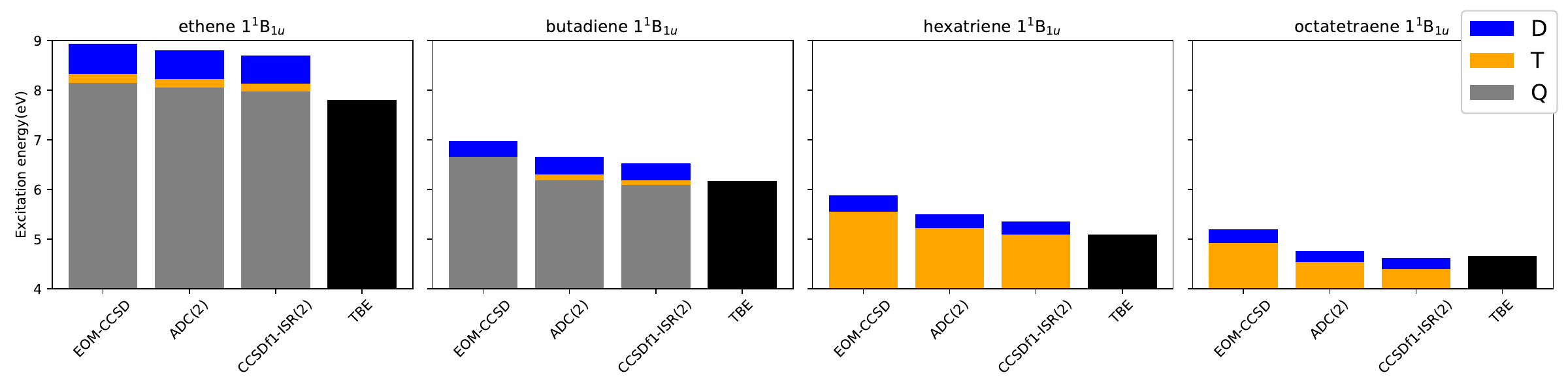}
    \caption{
    Basis set dependence of alkenes' 1$^1$B$_{1u}$ excitation energies. TBE data from Ref.~\citen{SchSilSau08}}
    \label{fig:polyenes_B1u}
\end{figure}
\begin{figure}[h!!]
    \centering
    \fig{0.45}{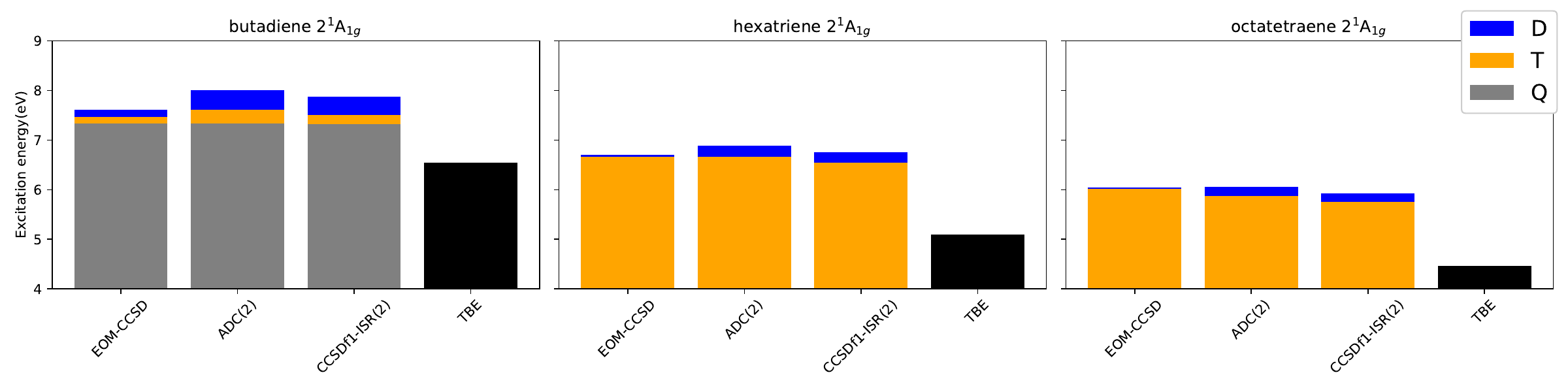}
    \caption{
    Basis set dependence of alkenes' 2$^1$A$_{1g}$ excitation energies. TBE data from Ref.~\citen{SchSilSau08}}
    \label{fig:polyenes_A1g}
\end{figure}

\newpage
\section{\textcolor{red}{N$_2$ spectroscopic parameters}}
\begin{table}[h]
\caption{N$_2$ ground state equilibrium bond length (\r{A}) and vibrational frequency (\text{cm}$^{-1}$)}
\label{tab:spec_param}
\resizebox{0.3\textwidth}{!}{
\begin{tabular}{|l|l|l|}
\hline
method & R$_{eq}$ & $\omega$ \\ \hline
CCSD   & 1.093  & 2449 \\ \hline
CCSDf1 & 1.095  & 2432   \\ \hline
CCD    & 1.092  & 2460   \\ \hline
CCDf1  & 1.094  & 2445   \\ \hline
CCD0   & 1.095  & 2423   \\ \hline
CCDf0 & 1.095  & 2425  \\ \hline
CCD1   & 1.073  & 2627   \\ \hline
pCCD   & 1.064  & 2715       \\ \hline
\end{tabular}
}
\end{table}
For our finite difference Hessian calculation on top of PySCF's CC analytical nuclear gradients, we took a step size $\Delta r = 0.0001 $\r{A}.

\newpage 
\section{\textcolor{red}{Equilibrium geometry N$_2$ and CO excitation energies from pCCD-based methods}}

\begin{table}[h]
\caption{N$_2$ and CO excitation energies (eV) in the aug-cc-pVTZ basis}
\label{tab:degen_pccd}
\begin{tabular}{c|c|l|l|l|c}
\hline\hline
\mc{1}{c}{system}   & \mc{1}{c}{state} & \mc{1}{c}{ADC(2)}	& \mc{1}{c}{EOM-pCCD+S}	& \mc{1}{c}{pCCD-ISR(2)} & \mc{1}{c}{TBE$^a$} \\ 
   \hline
 \multirow{6}{*}{CO} & $1\:^{1}\mathrm{\Sigma^+} \rightarrow 1\:^{1}\mathrm{\Pi}  (n \rightarrow \pi^\star)$ & 8.68& 9.355
& 6.358
&8.49\\  
 & $1\:^{1}\mathrm{\Sigma^+} \rightarrow 1\:^{1}\mathrm{\Sigma^-}  (\pi \rightarrow \pi^\star)$ &10.02&9.922&6.884&9.92\\
& $1\:^{1}\mathrm{\Sigma^+} \rightarrow 1\:^{1}\mathrm{\Delta}  (\pi \rightarrow \pi^\star)$ &10.30&9.943/10.354&7.151/7.298&10.06\\
& $1\:^{1}\mathrm{\Sigma^+} \rightarrow 2\:^{1}\mathrm{\Sigma^+}  $ &11.33&12.299&9.896&10.95\\
&$1\:^{1}\mathrm{\Sigma^+} \rightarrow 3\:^{1}\mathrm{\Sigma^+}  $ &11.84&12.866&10.497&11.52\\
& $1\:^{1}\mathrm{\Sigma^+} \rightarrow 2\:^{1}\mathrm{\Pi}  $ &12.04&13.097&10.671&11.72\\
    \hline
    
\multirow{7}{*}{N$_2$} & $1\:^{1}\mathrm{\Sigma_g} \rightarrow 1\:^{1}\mathrm{\Pi_g}  (n \rightarrow \pi^\star)$ & 9.47& 9.558/9.647& 
6.743&  9.34
\\
&$1\:^{1}\mathrm{\Sigma_g} \rightarrow 1\:^{1}\mathrm{\Sigma_u^-}  (\pi \rightarrow \pi^\star)$ &10.25&9.000&6.720&9.88\\
&$1\:^{1}\mathrm{\Sigma_g} \rightarrow 1\:^{1}\mathrm{\Delta_u}  (\pi \rightarrow \pi^\star)$ &10.79&10.399&7.221/7.489&10.29\\
&$1\:^{1}\mathrm{\Sigma_g} \rightarrow 1\:^{1}\mathrm{\Sigma_g^+}  $ &13.01&--&11.480&12.98\\
&$1\:^{1}\mathrm{\Sigma_g} \rightarrow 1\:^{1}\mathrm{\Pi_u}  $ &13.34&--&--&13.03\\
&$1\:^{1}\mathrm{\Sigma_g} \rightarrow 1\:^{1}\mathrm{\Sigma_u^+}  $ &13.08&--&--&13.09\\
&$1\:^{1}\mathrm{\Sigma_g} \rightarrow 2\:^{1}\mathrm{\Pi_u}  $ &13.95&--&--&13.46\\
   \hline\hline
   \mc{6}{l}{\fns $^a$Theoretical best estimates (TBE) and geometries from Ref.~\citen{LooSceBlo18}}
\end{tabular}
\end{table}

\section*{References}

\providecommand{\refin}[1]{\\ \textbf{Referenced in:} #1}